\documentclass[letterpaper]{article} 
\usepackage{aaai2026}  
\usepackage{times}  
\usepackage{helvet}  
\usepackage{courier}  
\usepackage[hyphens]{url}  
\usepackage{graphicx} 
\usepackage{natbib}  
\usepackage{caption} 
\DeclareCaptionStyle{ruled}{labelfont=normalfont,labelsep=colon,strut=off} 
\usepackage{algorithm}
\usepackage{algorithmic}
\usepackage{color, colortbl}
\usepackage[table]{xcolor}
\usepackage{url}
\usepackage{subcaption}
\usepackage{textcomp}
\usepackage{soul}
\usepackage{multirow}
\usepackage{enumitem}
\usepackage{mathtools}
\usepackage{siunitx}

\usepackage{booktabs} 
\usepackage{longtable,booktabs}

\usepackage{arydshln}
\definecolor{linkColor}{RGB}{6,125,233}
\definecolor{green}{rgb}{0.0, 0.65, 0.31}
\definecolor{bleudefrance}{rgb}{0.19, 0.55, 0.91}
\definecolor{ceruleanblue}{rgb}{0.16, 0.32, 0.75}
\definecolor{grey}{HTML}{969696}
\definecolor{violet}{HTML}{756bb1}
\definecolor{dgrey}{HTML}{01665e}
\definecolor{lgrey}{HTML}{5ab4ac}
\definecolor{dgreen}{HTML}{005a32}
\definecolor{purple}{HTML}{ae017e}

\definecolor{editCol}{HTML}{000000}
\definecolor{maskCol}{HTML}{c51b7d}
\definecolor{lrColor}{HTML}{8856a7}
\definecolor{trColor}{HTML}{d01c8b}
\definecolor{ctColor}{HTML}{4dac26}
\definecolor{brickred}{HTML}{f03b20}

\definecolor{sigone}{RGB}{222, 235, 247}    
\definecolor{sigtwo}{RGB}{158, 202, 225}    
\definecolor{sigthree}{RGB}{107, 174, 214}  

\definecolor{improveCol}{HTML}{4dac26}
\definecolor{worsenCol}{HTML}{d01c8b}

\definecolor{DarkBlue}{HTML}{00008B}
\definecolor{mscolor}{HTML}{01665e}
\definecolor{nmscolor}{HTML}{bf812d}
\definecolor{lgreen}{HTML}{ccece6}
\definecolor{dolive}{HTML}{308014}

\colorlet{tablerowcolor4}{gray!50} 

\newcommand*{\textlabel}[2]{%
  \edef\@currentlabel{#1}
  \phantomsection
  #1\label{#2}
}

\colorlet{tableheadcolor}{gray!25} 
\colorlet{tablerowcolor}{gray!10} 
\colorlet{tablerowcolor2}{gray!45} 
\colorlet{tablerowcolor3}{gray!12} 

\newcommand{\rowcollight}{\rowcolor{tablerowcolor3}} %

\newcolumntype{a}{>{\columncolor{tablerowcolor}}r}
\definecolor{aicolor}{HTML}{018571}
\definecolor{occolor}{HTML}{ff7799}

\definecolor{aicolor}{HTML}{fc8d62}
\definecolor{occolor}{HTML}{253494}

\newif{\ifhidecomments}
  \hidecommentsfalse 
\ifhidecomments
    \newcommand{\olivia}[1]{}
    \newcommand{\agam}[1]{}
    \newcommand{\eshwar}[1]{}
    \newcommand{\koustuv}[1]{}
\else
    \newcommand{\olivia}[1]{\textbf{\small\sffamily{\textcolor{DarkBlue}{[#1 -- Olivia]}}}}
    \newcommand{\agam}[1]{\textbf{\small\sffamily{\textcolor{dgreen}{[#1 -- Agam]}}}}
    \newcommand{\eshwar}[1]{\textbf{\small\sffamily{\textcolor{orange}{[#1 -- Eshwar]}}}}
    \newcommand{\koustuv}[1]{\textbf{\small\sffamily{\textcolor{purple}{[#1 -- Koustuv]}}}}
  \fi

\newcommand{\Tr}{\textit{Treated}}
\newcommand{\Ct}{\textit{Control}}

\newcommand{\bsky}{\textit{Bluesky}}
\newcommand{\news}{\texttt{News}}

\colorlet{tableheadcolor}{gray!25} 

\definecolor{neutralCol}{HTML}{dd1c77}
\definecolor{neutralGreen}{HTML}{31a354}
\definecolor{NewBlue}{HTML}{1879ba}
\definecolor{bleudefrance}{rgb}{0.19, 0.55, 0.91}  
\definecolor{AfTrColor}{HTML}{0868ac}  
\definecolor{BfTrColor}{HTML}{a8ddb5}  

\definecolor{AfCtColor}{HTML}{b10026}  
\definecolor{BfCtColor}{HTML}{fd8d3c}

\graphicspath{ {figures/} }

\newcommand{\para}[1]{\vspace{0.2em}\noindent\textbf{\textit{#1}~}}

\usepackage{xcolor}

\usepackage{newfloat}
\usepackage{listings}
\floatstyle{ruled}
\newfloat{listing}{tb}{lst}{}
\floatname{listing}{Listing}
\title{The Hidden Toll of Social Media News: Causal Effects on Psychosocial Wellbeing}
\author {
    Olivia Pal, Agam Goyal, Eshwar Chandrasekharan, Koustuv Saha 
}
\affiliations {
    University of Illinois Urbana-Champaign\\
    Urbana, IL, USA\\
    \{opal2, agam2, eshwar, ksaha2\}@illinois.edu
}

\usepackage{bibentry}

\begin{document}

\maketitle

\begin{abstract}


News consumption on social media has become ubiquitous, yet how different forms of engagement shape psychosocial outcomes remains unclear. 
To address this gap, we leveraged a large-scale dataset of $\sim$26M posts and $\sim$45M comments on the \bsky{} platform, and conducted a quasi-experimental study, matching 81,345 \Tr{} users exposed to \news{} feeds with 83,711 \Ct{} users using stratified propensity score analysis. We examined psychosocial wellbeing, in terms of \textit{affective}, \textit{behavioral}, and \textit{cognitive} outcomes. 
Our findings reveal that news engagement produces systematic trade-offs: increased depression, stress, and anxiety, yet decreased loneliness and increased social interaction on the platform. 
Regression models reveal that \news{} feed bookmarking is associated with greater psychosocial deterioration compared to commenting or quoting, with magnitude differences exceeding tenfold. 
These per-engagement effects accumulate with repeated exposure, showing significant psychosocial impacts. Our work extends theories of news effects beyond crisis-centric frameworks by demonstrating that routine consumption creates distinct psychological dynamics depending on engagement type, and bears implications for tools and interventions for mitigating the psychosocial costs of news consumption on social media.

\end{abstract}

\section{Introduction~\label{section:intro}}




\begin{quote}
\small
\textit{Psychologists are seeing an increase in news-related stress.}---American Psychological Association~\cite{huff2022media}
\end{quote}

News occupies a distinct position within social media ecosystems, serving as an emotionally salient and socially consequential form of content that is encountered routinely and often incidentally as part of everyday online activity.
News consumption has become a central form of engagement on social media platforms. 
Rather than occurring during isolated moments of intentional information seeking, news exposure now unfolds as part of everyday online activity. 
In 2025, more than half of U.S. adults (53\%) reported that they at least sometimes get news from social media~\cite{pewsocialmedia2025}.
These platforms further transform news consumption by enabling users to engage with news content in interactive ways through liking, commenting, and sharing news content~\cite{hermida2010twittering}. 
As a result, users do not merely consume news, but actively shape their informational environments through repeated and varied forms of engagement. This shift raises important questions about how different modes and intensities of news engagement affect users' wellbeing.

In parallel, prior work has studied the psychosocial effects of news exposure---reflected in changes to emotional states, stress-related behaviors, and cognitive expressions---primarily, in the context of crises.
For example, a survey of 2,251 adults found that more frequent information seeking about COVID-19 across television, newspapers, and social media was associated with higher levels of emotional distress~\cite{hwang2021relationship}.
Similar patterns have been observed following mass shootings, terrorist attacks, and natural disasters
\cite{silver2013mental,strasser2022covid}.~\citeauthor{thompson2019media}
further showed that psychological distress can drive increased news consumption, reinforcing negative outcomes~\cite{thompson2019media}.

Notably, crisis-driven events account for only a fraction of the news users encounter online; everyday social media use involves ongoing exposure to political, economic, and local news~\cite{boczkowski2018news}. 
Such sustained news consumption can contribute to emotional exhaustion and reduced wellbeing, with news avoidance emerging as a common coping response~\cite{mclaughlin2023caught,soroka2015news,boukes2017news}. 
These experiences are often described in popular discourse as headline anxiety'' or headline stress disorder,'' which may be more pronounced on social media where news is encountered repeatedly and incidentally within everyday feeds~\cite{huff2022media}.

Understanding the effects of social media news engagement is critical, as routine and repeated exposure may accumulate psychological effects distinct from crisis-driven media exposure and general social media use~\cite{silver2013mental,soroka2015news}. 
Isolating these effects is essential for advancing media-effects theory and informing platform design and policy interventions aimed at mitigating unintended psychological consequences~\cite{thorson2016curated,newman2025digital,huff2022media}.

Existing research on news consumption and mental health has largely relied on survey-based studies and laboratory experiments~\cite{boukes2017news,strasser2022covid}. While informative, these approaches are limited in social media contexts: surveys are retrospective and susceptible to recall and subjective bias~\cite{tourangeau2000psychology}, and experimental designs often capture short-term responses under controlled conditions that may not reflect everyday patterns of news engagement online.
Consequently, it remains unclear how news engagement shapes psychosocial outcomes over extended periods across large populations in naturalistic settings, and whether these effects vary by the type and intensity of engagement. 
This challenge is compounded by the difficulty of observing users early in their engagement trajectories, as most platforms capture individuals only after years of accumulated habits and algorithmic personalization~\cite{thorson2016curated}.
Without such longitudinal examinations, existing studies risk conflating the effects of news exposure with broader platform use, obscuring the mechanisms through which specific engagement modalities influence wellbeing.

To address these gaps, we conducted a large-scale quasi-experimental study of news engagement on \bsky{}---a decentralized (Twitter-like) social media platform that was launched publicly in February 2024. 
\bsky{} provides a unique opportunity to examine news engagement from its early stages, allowing observation of users as they develop engagement practices from the ground up~\cite{failla2024m}. Unlike established platforms where algorithmic news exposure is often implicit, \bsky{} users explicitly choose whether to bookmark algorithmic feeds, including a verified \news{} feed that aggregates headlines from established news organizations. 
This paper is guided by the following research questions (RQs):

\begin{enumerate}[align=left]
\item[\textbf{RQ1:}] How does news exposure on social media affect psychosocial wellbeing?
\item[\textbf{RQ2:}] How are the types and intensity of news engagement on social media associated with psychosocial outcomes?
\end{enumerate}

To address our RQs, we adopted a causal-inference design, based on the potential outcomes framework~\cite{rubin2005causal}, on a large-scale dataset on \bsky{}.
We identified a \Tr{} group consisting of users who engaged with the \news{} feed through bookmarking, liking, commenting, reposting, quoting, or creating original posts. 
We compared these users to stratified propensity-score-matched \Ct{} users who were active on the platform but never engaged with news content. 
We examined psychosocial outcomes across \textit{affective}, \textit{behavioral}, and \textit{cognitive} dimensions.


Our findings reveal a complex pattern of effects. 
News exposure is associated with increases in depression, stress, and anxiety, yet simultaneously linked to decreased loneliness and increased social participation, with \Tr{} users commenting and quoting at substantially higher rates than \Ct{} users.
These results highlight a fundamental trade-off: \textit{everyday news engagement can enable social connection, but also strain emotional wellbeing.}
Importantly, not all engagement forms carry equal risks: 
the quantity of
bookmarking, has 1900\% greater psychosocial effects than other forms of engagement (e.g., commenting, quoting).

Together, these findings underscore the need to move beyond binary assessments of whether news exposure is ``good'' or ``bad'' for mental health. 
Our study highlights the importance of engagement forms as a key design and behavioral lever, and bears implications for platform design and user behavior. 
For instance, platform design that shifts users from accumulated feed consumption toward more social engagement may enable users to draw benefits of news consumption, while limiting the psychosocial costs. 

\section{Related Work}\label{section:rw}


\para{Effects of Media and News Consumption.}
News consumption has long been a central focus of media-effects research. 
Early research on agenda-setting demonstrated that news coverage shapes what issues people attend to, rather than what opinions they hold about those issues~\cite{mccombs1972agenda}. 
Related research on cultivation effects showed that repeated exposure to television news---particularly violent crime coverage---can lead to heightened perceptions of risk and fear even when objective threat remains low~\cite{gerbner1980mainstreaming}. 
These effects emerged in media environments characterized by editorial control, scheduled broadcasts, and largely one-way communication.

Over the years, the rise of the internet and social media has fundamentally altered how people encounter news. 
Today, over half of U.S. adults report getting news from social media platforms, with sites such as Facebook and YouTube serving as major news sources~\cite{mitchellpew}. 
Unlike traditional media, social media enables algorithmic curation, personalized feeds, and real-time interaction, allowing users to encounter, share, and respond to news continuously as part of everyday online activity. 
This shift blurs the boundary between intentional news seeking and incidental exposure, transforming news consumption from a discrete activity into an ambient experience.

Research on the psychological effects of online news exposure has largely focused on crisis contexts. 
Studies of the COVID-19 pandemic, mass shootings, terrorist attacks, and natural disasters link heightened exposure to distressing news with increased stress, anxiety, and emotional exhaustion~\cite{silver2013mental,strasser2022covid,thompson2019media}. Public discourse has similarly described these experiences as forms of ``headline anxiety'' or ``headline stress,'' reflecting the cognitive and emotional strain associated with repeated exposure to alarming news~\cite{huff2022media}. Such effects may be especially pronounced on social media, where headlines are encountered frequently and incidentally within routine feeds.

Everyday social media use involves ongoing exposure to several types of news, which prior work suggests can also contribute to emotional exhaustion and reduced wellbeing, with news avoidance emerging as a common coping response~\cite{mclaughlin2023caught,soroka2015news,boukes2017news}. 
However, the psychosocial effects of repeated and incidental news exposure on social media remain understudied; our work aims to address this gap.

\para{Social Media and Mental Health.}
Prior work has shown how social media platforms enable candid self-disclosure of mental health experiences~\cite{de2014mental}, and how behavioral and linguistic signals on social media can provide meaningful insights into mental health states~\cite{de2016discovering}. 
Studies on platforms such as Twitter have demonstrated that changes in posting behavior, emotional expression, and self-focused language can precede clinical depression diagnoses by ~\cite{de2013predicting,coppersmith2014quantifying}. 
Accordingly, social media data has enabled large-scale study of psychological responses to external events, including crises~\cite{palen2008online,saha2017stress}, wars and conflicts~\cite{mark2012blogs,de2014narco}, natural disasters~\cite{starbird2010chatter}, crime and violence~\cite{valdes2015psychological}, and terrorist attacks~\cite{cohn2004linguistic,lin2014ripple}. 
Across these contexts, social media platforms often function as spaces for self-expression, support, and solidarity during periods of crisis~\cite{mark2012blogs,starbird2010chatter}.

This line of work has adopted computational frameworks that operationalize psychological constructs from social media language. 
The Linguistic Inquiry and Word Count (LIWC) framework has been widely used to capture psycholinguistics~\cite{tausczik2010psychological}. 
Prior work has also developed classifiers to identify symptomatic mental health expressions, including stress, anxiety, and suicidal ideation~\cite{saha2019social,lin2014user,burnap2015machine}, and shown the construct validity of these approaches~\cite{saha2022social,ernala2019methodological}. Building on these advances, we leverage established linguistic and behavioral measures to examine the psychosocial effects of everyday news consumption.

\para{Causal Inference Approaches on Social Media and Observational Data.}
Ideally, causal inference is established through randomized trials and experimental designs. However, ethical considerations and practical constraints often limit their feasibility.
In observational data settings, such as on longitudinal and large-scale social media data, recent research has drawn on the potential outcomes framework to estimate causal effects by mitigating confounds~\cite{rubin2005causal}.
Therefore, matching-based quasi-experimental approaches have been used to construct comparable treatment and control groups in social media studies. 
Prior work has adopted these approaches for online platform-based interventions~\cite{saha2021advertiming,jhaver2024bystanders,chandrasekharan2017you,chowdhury2021examining,lambert2025does,chan2025examining}.
Prior work has leveraged these approaches to examine a range of interventions, including college alcohol use~\cite{kiciman2018using}, psychiatric medication use~\cite{saha2019social}, crisis events~\cite{saha2017stress}, public service interventions~\cite{saha2018social}, online support~\cite{de2017language,saha2020causal}, and shifts in suicidal ideation~\cite{de2016discovering}. 
Related studies have examined mental health coping narratives~\cite{yuan2023mental}, county-level health estimates using Twitter data~\cite{culotta2014estimating}, and the mental health impacts of the COVID-19 pandemic~\cite{saha2020psychosocial,jha2021learning,vowels2023toward}.

Methodological advances have enabled the estimation of heterogeneous treatment effects, allowing causal effects to vary across users rather than assuming uniform impact~\cite{xie2012estimating,olteanu2019social,saha2021advertiming}. 
Prior work has examined how exposure to ideologically diverse news shapes political attitudes~\cite{bakshy2015exposure}, how misinformation exposure influences beliefs and behavior~\cite{guess2020digital}, how social endorsements affect news selection~\cite{messing2014selective}, and how exposure to hateful speech affects stress expressions~\cite{saha2019prevalence}. 
Building on this literature, our work adopts a quasi-experimental approach to estimate the psychosocial effects of news engagement on social media.

\section{Data}\label{sec:data}
We study the psychosocial effects of news engagement on social media, particularly on the \bsky{} platform.
\bsky{} is a decentralized, Twitter-like social media platform that was initially launched as an invite-only service and opened to the public in February 2024~\cite{failla2024m}. Prior work characterizes \bsky{} as exhibiting an activity distribution comparable to that of other social media platforms, while also reporting a higher proportion of original content relative to reshared posts and comparatively lower levels of toxicity~\cite{nogara2025}.




\begin{table}[t]
\centering
\sffamily
\small
\resizebox{\columnwidth}{!}{
\begin{tabular}{lrrrr}
\textbf{Dataset} $\rightarrow$ & \multicolumn{2}{c}{\textbf{\Tr{}}} & \multicolumn{2}{c}{\textbf{\Ct{}}} \\
\cmidrule(lr){1-1}\cmidrule(lr){2-3}\cmidrule(lr){4-5}
\textbf{Metric} $\downarrow$ & \textbf{Count} & \textbf{Mean} & \textbf{Count} & \textbf{Mean}\\
\toprule
Users & 148,680 & -- & 2,221,748 & -- \\
Posts & 19,245,131 & 225.02 & 7,227,387 & 56.64\\
Quotes & 8,102,224 & 94.73 & 540,409 & 4.23\\
\hdashline
Comments & 39,370,440 & 460.33 & 6,303,635 & 49.40\\
Reposts & 33,467,258 & 391.31 & 4,181,332 & 32.77\\
Likes & 4,034,615 & 47.17 & 54,854 & 0.43\\
\end{tabular}}
\caption{Descriptive Statistics of the entire \bsky{} dataset.}
\label{tab:descriptives}
\end{table}

We obtained our dataset from the \bsky{} data collated by prior work~\cite{failla2024m}, which contains data from over 80\% of all registered accounts at the time of collection.
The \bsky{} platform consists of feed generators, which are essentially content recommendation algorithms. 
These feeds can be subscribed by users based on their preferences and interests. 
This dataset had the data of 11 such thematic feeds, which were available on the platform. 


We adopted a quasi-experimental study design based on the potential outcomes framework~\cite{rubin2005causal}, where the \textit{treatment} is operationalized as the exposure to the \news{} feed. 
The \news{} feed is a verified feed generator on \bsky{} that aggregates headlines from verified news organizations, with the most recent content appearing first~\cite{nogara2025}. 
This feed provides users with a centralized stream of news content covering current events, politics, and public affairs from established journalistic sources.

We compiled two groups: a \Tr{} group of users who engaged with the \news{} feed, and a \Ct{} group of users who never engaged with it. 
For each user, we obtained the entire longitudinal data of all interactions, including bookmarking, posting, liking, commenting, reposting, and quoting posts.
We describe our approach below:


\subsubsection{Compiling \Tr{} Dataset}
For our study, we define treatment as the exposure to the \news{} feed. 
We define \Tr{} users to be a group of users who were exposed to the \news{} feed, where exposure could be in the form of \textit{bookmarking}, \textit{liking}, \textit{commenting}, \textit{reposting}, \textit{quoting}, or \textit{posting} on this feed. 
We gathered the longitudinal timeline of all the users who integrated feed-specific post listings, feed bookmarks, post-level likes, and users' posting histories as well as users who indirectly interacted with the \news{} feed through commenting, reposting, and quoting others' posts that appeared on the \news{} feed.
Finally, we obtained a total of 148,680 \Tr{} users, and recorded the date of their first exposure to the \news{} feed as the treatment date. 


\subsubsection{Compiling \Ct{} Dataset}
\begin{figure*}[t]
\centering
\begin{subfigure}[b]{0.77\columnwidth}
    \centering
    \includegraphics[width=\columnwidth]{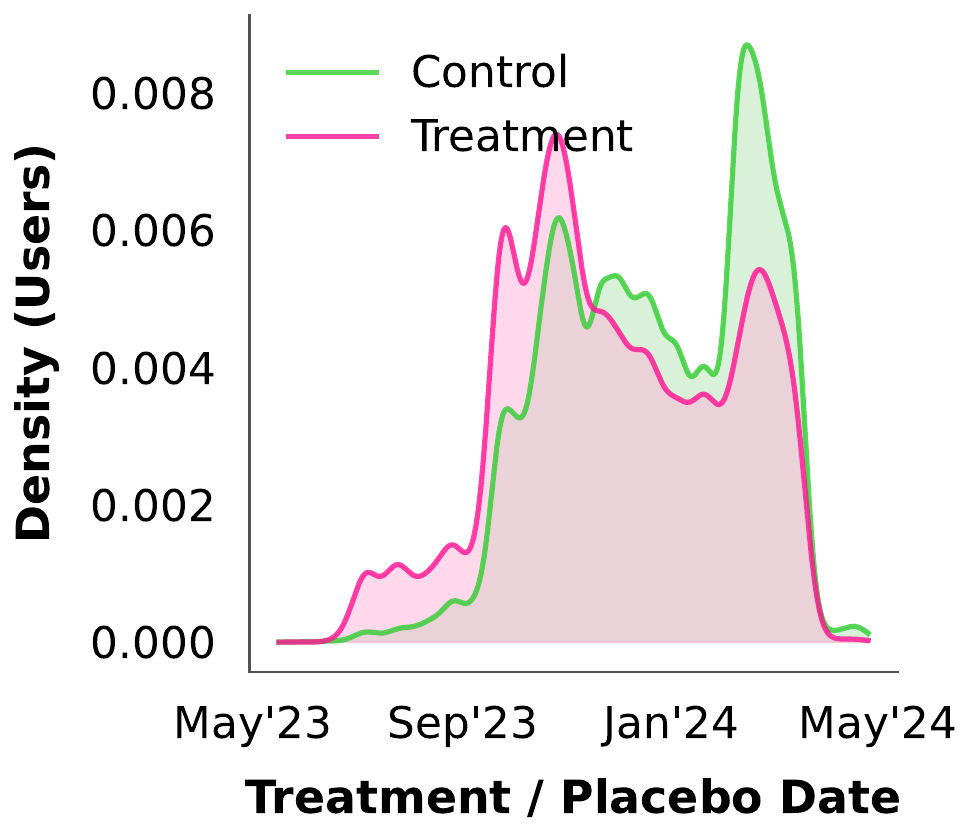}
    \caption{}
    \label{fig:anchor_dates}
    \end{subfigure}\hfill
    \begin{subfigure}[b]{0.66\columnwidth}
    \centering
    \includegraphics[width=\columnwidth]{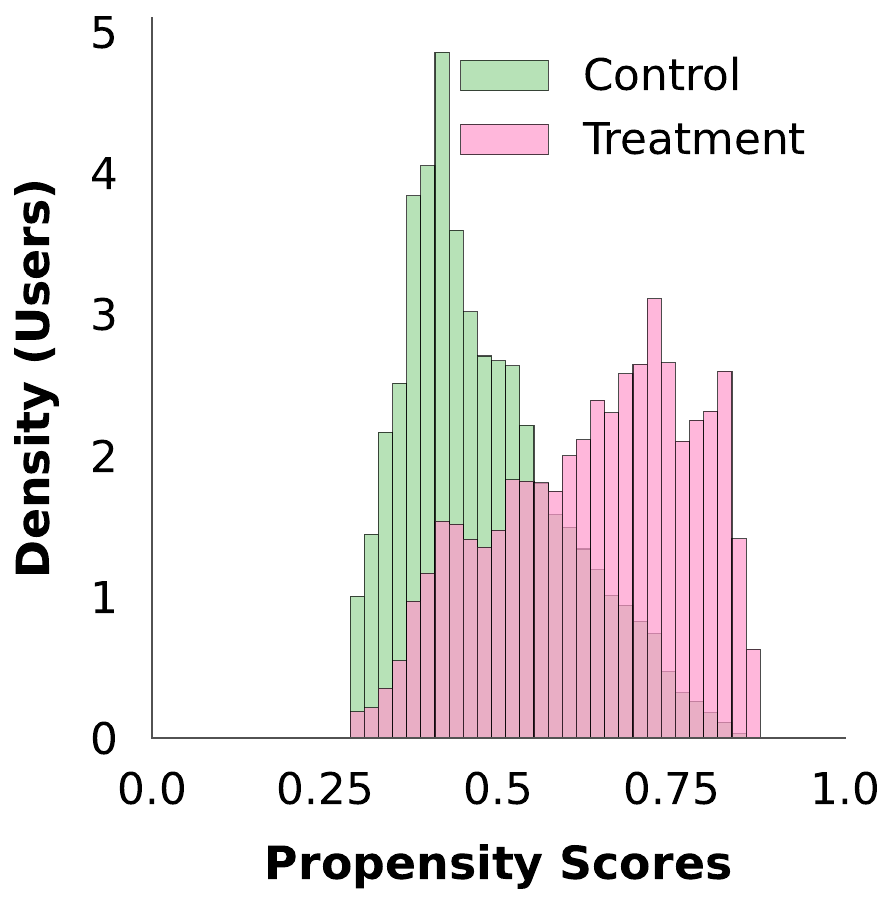}
    \caption{}
    \label{fig:propensity_score}
    \end{subfigure}\hfill
\begin{subfigure}[b]{0.66\columnwidth}
    \centering
    \includegraphics[width=\columnwidth]{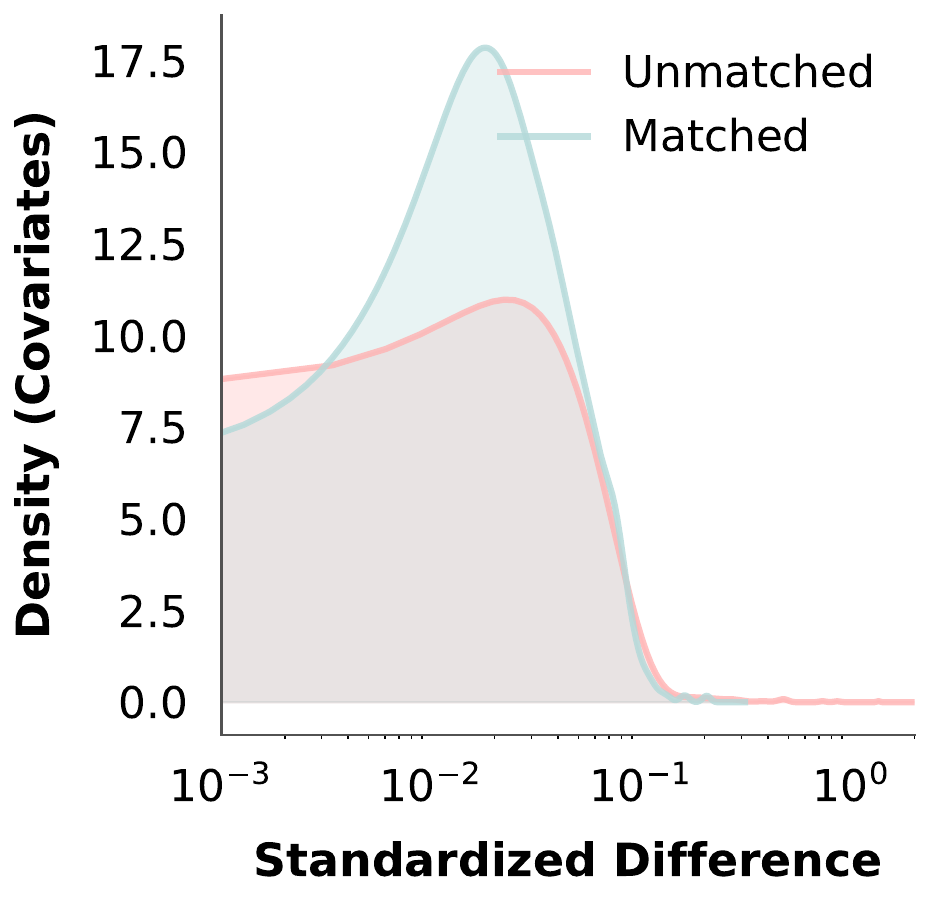}
    \caption{}
    \label{fig:smd-density}
    \end{subfigure}\hfill
\caption{Distribution of a) treatment/placebo dates, b) propensity scores, and b) standardized mean differences (SMD) across the \Tr{} and \Ct{} users' datasets.}
\end{figure*}

We built a \Ct{} dataset consisting of \bsky{} users who were exposed to algorithmic feeds, but never engaged with the \news{} feed.
Similar to the \Tr{} users, we compiled the entire longitudinal timeline of these users. 
Given that the \Ct{} users never engaged with the \news{} feed, their timelines did not have an actual treatment date.
Therefore, for comparability with the \Tr{} users, we assigned placebo dates. 
We assigned each \Ct{} user a placebo date, matching the non-parametric distribution of treatment dates of the \Tr{} dataset,
to mitigate the effects of any temporal confounds.
For this, we ensured that the treatment and placebo dates follow similar distribution by non-parametrically simulating placebo dates from the pool of treatment dates. 
We measured the similarity in the distribution of treatment and placebo dates using Kolmogorov--Smirnov test
to obtain an extremely low statistic of 0.06, indicating similarity in the probability distribution of treatment and placebo dates (Figure~\ref{fig:anchor_dates}).

Table~\ref{tab:descriptives} shows the descriptive statistics of our entire dataset, consisting of 2.37M users, 26M posts, and 45M comments, as a starting point for our ensuing analyses.

\section{Methods}
To study the psychosocial effects of being exposed to \news{} on \bsky{}, we adopted a causal inference approach~\cite{rubin2005causal}.
Ideally, causal inference studies are best conducted with randomized controlled trials (RCTs), which eliminate selection bias by randomly assigning participants to treatment and control conditions. 
However, RCTs are often impractical and can raise ethical concerns when investigating naturally occurring social media exposures~\cite{small2024protocols, moreno2013ethics, grimmelmann2015law}. 
Therefore, in this work we adopt a quasi-experimental design, based on the potential outcomes framework~\cite{rubin2005causal} which helps simulate a randomized controlled trial (RCT) with observational data to examine whether an outcome is caused by a treatment $T$, by comparing two potential outcomes: 1) $Y_i (T=1)$ when exposed to $T$, and 2) $Y_i (T=0)$ if there was no $T$. 
However, it is impossible to obtain both of these outcomes for the same individual. 
To overcome this challenge of missing data, this framework estimates the missing counterfactual outcome for an individual based on the outcomes of other similar (matched) individuals.
In particular, we employed stratified propensity score analysis to match and then to examine the psychosocial outcomes in \Tr{} and \Ct{} individuals by measuring the relative treatment effect (RTE) of news exposure~\cite{saha2019social,kiciman2018using}. 


\begin{figure}[t]
\centering
\includegraphics[width=\columnwidth]{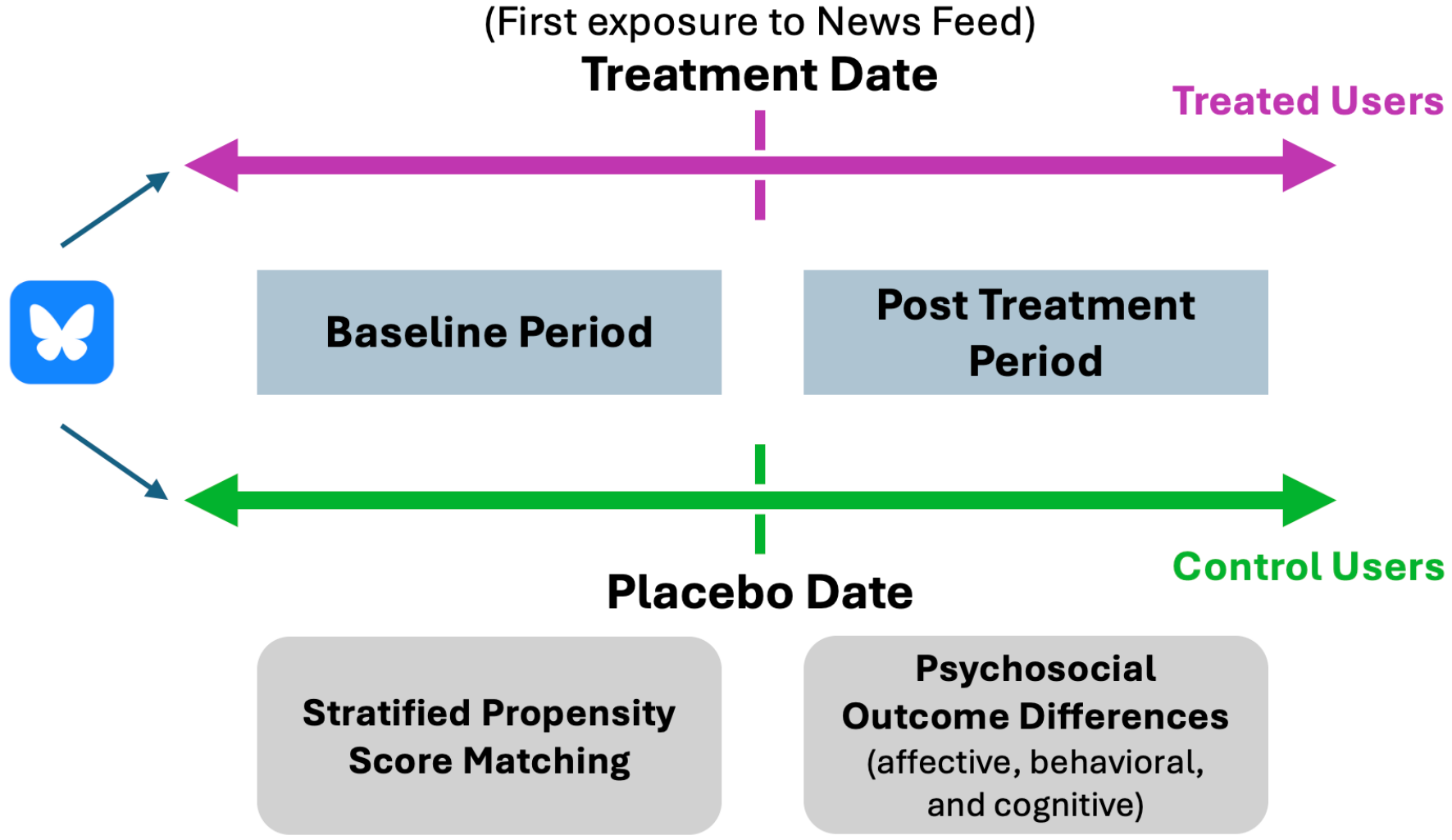}
\caption{A schematic figure showing our causal-inference approach to analyze users' \bsky{} timeline.}
\label{fig:schematic}
\end{figure}

For this purpose, we estimated each user's likelihood, i.e., propensity score of encountering news content, using an AdaBoost classifier trained on baseline covariates. 
Users are then stratified into propensity score bins, creating matched groups where \Tr{} users (those exposed to news) and \Ct{} users (those not exposed).
Within each matched stratum of the \Tr{} and \Ct{} groups, we compared post-treatment psychosocial outcomes.
This section elaborates our approach (see Figure~\ref{fig:schematic} for a schematic overview).

\subsection{Operationalizing Psychosocial Outcomes}

To measure the psychosocial effects of being exposed to the \news{} feed on \bsky{}, we draw from the tripartite model of psychological wellbeing ~\cite{breckler1984empirical}, which states that psychological states comprise three distinct components: \textit{affective} (emotional experiences), \textit{behavioral} (overt actions and social engagement), and \textit{cognitive} (thoughts and mental processes). 
Inspired by this framework and prior research at the intersection of social media and mental health~\cite{saha2018social, saha2020causal, yuan2023mental}, we operationalized these categories as psychosocial outcomes, explained below:

\subsubsection{Affective Outcomes}
Affective outcomes capture individuals' emotional experiences and psychological states. 
We measured two types of affective outcomes as below:

\para{Symptomatic Mental Health Expressions.} We obtained the symptomatic mental health expressions of \textit{depression}, \textit{anxiety}, \textit{stress}, \textit{suicidal ideation}, and loneliness in users' language using SVM classifiers from Saha et al.~\cite{saha2019social}. 
Each classifier uses $n$-gram ($n$=1,2,3) features and was trained on posts from specific mental health communities on Reddit. The training data includes posts from r/depression, r/anxiety, r/stress,  r/SuicideWatch, and r/lonely, with non-mental-health posts serving as negative examples. These models achieve approximately 90\% accuracy and have proved reliability across various platforms~\cite{saha2019social,saha2020causal,yuan2025mental}.
For each user and time period, we concatenated all their posts and applied each classifier to 
obtain a normalized distribution of these symptomatic expressions.


\para{Psycholinguistic affect.} We extracted affect word frequencies using LIWC~\cite{pennebaker2015development}. 
We counted words in the affect categories (positive emotion, negative emotion, anxiety, anger, and sadness), normalizing by post length (per 100 words). 

\subsubsection{Behavioral Outcomes}
Behavioral outcomes include an individual's overt actions, behavioral intentions, and verbal statements regarding behavior~\cite{breckler1984empirical}. 
Prior studies quantified behavioral psychological wellbeing by measuring the shifts in social functioning and interests~\cite{guntuku2019language, saha2018social}, and we obtained the following measures as behavioral outcomes:

\para{Activity.} To investigate whether news exposure shifts an individual's \bsky{} posting activity, 
we calculated the average number of posts per day, quotes per day, and comments per day for every individual. We also calculated authored rows per day, which captures the total self-authored content each user has per day.

\para{Interactivity.} Interactivity is another indicator of an individual showing psychosocial changes~\cite{saha2020causal}. We measured interactivity based on a user's \bsky{} participation as a ratio of interactive posts (replies and quotes) to self-authored posts. 


\subsubsection{Cognitive Outcomes}
Cognitive outcomes reflect an individual's thought patterns, mental processes, and information processing capabilities~\cite{breckler1984empirical}. 
Drawing from psycholinguistic research demonstrating that language style and structure reveal cognitive functioning~\cite{pennebaker2003words, tausczik2010psychological}, we operationalized cognitive wellbeing through linguistic markers including verbosity, readability, complexity, and repeatability: 


\para{Readability.} Readability reflects how easily a reader can comprehend a given text. In mental health contexts, it plays a crucial role in both expression and interpretation ~\cite{ernala2017linguistic, park2018harnessing}. We measured readability using the Coleman-Liau Index (CLI), which calculates readability based on character and sentence counts: $\text{CLI} = 0.0588 \times L - 0.296 \times S - 15.8$ where $L$ represents average letters per 100 words and $S$ represents average sentences per 100 words. Higher readability would indicate 
a higher quality of writing. 

\para{Verbosity.} Verbosity is a measure of detail and elaboration in communication. We measured verbosity as the number of \textit{words per sentence}~\cite{saha2020causal}.

\para{Complexity and Repeatability.} Prior work has shown that higher levels of linguistic complexity and lower repetition correlate with better mental health~\cite{ernala2017linguistic}.
We measured \textit{repeatability} as the frequency of word reuse, where higher values may indicate lower communication quality due to redundancy. 
We measured \textit{complexity} as the average length of words per sentence, and greater complexity is often associated with more nuanced, precise, and detailed expression~\cite{kolden2011congruence}.


\para{Psycholinguistic Cognition and Social Contexts.} Using LIWC, we extracted word categories that reflect cognitive and social processes, grouped as: 1) \textit{Cognitive Processes} (causation, certainty, discrepancy, tentativeness, insight, negation), 2) \textit{Social Focus} (personal pronouns across all persons, impersonal references), and 3) \textit{Function Words} (articles, prepositions, auxiliary verbs, conjunctions, quantifiers, comparatives, and informal language markers).
Greater use of these linguistic patterns has been linked to better psychosocial wellbeing ~\cite{tausczik2010psychological, pennebaker2003words}.


\subsection{Matching for Causal Inference}



\subsubsection{Covariates}
When conditioned on high-dimensional covariate data, matching is known to mitigate bias compared to naive correlational analyses~\cite{rubin2005causal}. 
Our approach controls for covariates so that \Tr{} and \Ct{} groups show similar online behavior before being subjected to treatment (in our case, exposure to \news{} feed).
We draw on prior work to identify a range of covariates: 1) \textbf{activity metrics}: \emph{posting frequency} (number of posts per day) and \emph{posting tenure} (number of days on \bsky{}), 2)  \textbf{symptomatic mental health expressions:} classifier-based measures of \textit{depression, anxiety, stress, suicidal ideation}, and \textit{loneliness}~\cite{saha2019social,yuan2025mental}, 3) \textbf{content}: top 500 $n$-grams ($n$=2,3), and 4) \textbf{psycholinguistics}: all LIWC-22 attributes~\cite{pennebaker2001linguistic}.    
In total, we obtained 520 covariates and normalized the measures per user, based on each user's baseline period of a minimum of 30-days before the treatment date.

\subsubsection{Stratified Propensity Score Matching}
We used matching to find pairs (generalizable to groups) of \Tr{} and \Ct{} users whose covariates are statistically similar in the baseline period. 
In particular, we employed the propensity score matching (PSM), which matches users based on the likelihood of news exposure given baseline covariates~\cite{saha2019social}. 
For this purpose, we built an Adaboost classifier using the SAMME algorithm with decision trees (max depth=2) as base learners (rate=0.05, estinmators=200) (Zhu et al.~\citeyear{zhu2009multi}, Goyal et al.~\citeyear{goyal2025language}).
This model predicted the propensity scores ranging between 0 and 1.
Figure~\ref{fig:propensity_score} shows a distribution of propensity scores.



Next, we adopted a stratification approach---stratified matching enables to handle the bias-variance tradeoff by striking a balance between too biased (one-to-one matching) and too variant (unmatched) data comparisons~\cite{kiciman2018using}.
In particular, we stratified all the users into 10 equal-width strata based on their propensity scores, where each stratum consisted of users with similar baseline covariates, i.e., matched \Tr{} and \Ct{} users~\cite{saha2019social}. 

We filtered out users with no posting activity, users with baseline periods shorter than 30 days, and users with no post-treatment activity. 
Additionally, we used a minimum threshold of at least 10 users per group and dropped any strata with fewer users per group. 
Accordingly, we dropped three strata: the remaining seven strata, consisting of 83,711 \Tr{} users and 81,345 \Ct{} users, were used for our ensuing analysis. 



\subsubsection{Quality of Matching}
To ensure that our matching obtained comparable \Tr{} and \Ct{} users, we evaluated the balance of covariates. 
We obtained the standardized mean differences (SMD) for all the covariates across the \Tr{} and \Ct{} users. 
For each covariate, SMD is the difference in the mean values between the two groups as a fraction of the pooled standard deviation between the two groups.
Two groups can be considered to be balanced if the SMD is lower than 0.25~\cite{kiciman2018using}.
We find a significant reduction in the SMD from 0.042 (max=1.34) in the unmatched datasets to 0.037 (max=0.21) in the matched datasets (Figure \ref{fig:smd-density}).
Using a conservative threshold of 0.15 for adequate covariate balance, the number of imbalanced covariates decreased from 16 (3.1\%) before matching to 4 (0.8\%) after matching, reinforcing that the stratification successfully created comparable treatment and control groups.


\subsubsection{Relative Treatment Effect}
To measure the psychosocial effects of exposure to \news{}, we computed the Relative Treatment Effect (RTE) for each outcome~\cite{kiciman2018using}. 
We first calculated the RTE per outcome measure in every stratum, as a ratio of the mean outcome value in the \Tr{} group to that in the \Ct{} group. 
Next, using a weighted average across the strata, we obtained RTE per outcome measure. 
An outcome RTE greater than 1 would indicate that the outcome increased in \Tr{} users, whereas an RTE lower than 1 would indicate that it decreased in \Tr{} users, following the treatment (news exposure). 
We obtained effect size (Cohen's $d$) and conducted $t$-tests to assess statistical significance in differences.
Further, in order to examine the duration of these effects, we also plotted a temporal representation of the RTE over a two-month period following treatment.

\subsubsection{Individual Treatment Effect}
To quantify the difference in psychosocial effects across different forms of news engagement, we estimate heterogeneous treatment effects as a function of engagement types. 
Within each stratum, we computed the Individual Treatment Effect (ITE) for each user as the difference between their post-treatment outcome and the mean outcome of matched control users in the same stratum.
We then modeled these ITEs using ordinary least squares (OLS) regression without an intercept. 
Our engagement types are: 1) \textit{Posting} (original content creation), 2) \textit{Commenting} (direct reply to others' posts), 3) \textit{Quoting} (sharing someone else's post with additional comments), 4) \textit{Reposting} (sharing someone else's post with no added comments) 5)\textit{Bookmarking} (bookmarking a feed), and 6) \textit{Liking} (liking an existing post). 
For each outcome, we built a separate regression model, with the ITE as the dependent variable, and all the engagement types, along with baseline measures of mental health (depression, anxiety, stress, suicidal ideation, loneliness), linguistic features (readability, verbosity, repeatability), and interactivity were used as independent variables. 
To address multicollinearity among covariates, we excluded covariates with a variance inflation factor (VIF) exceeding 20 ~\cite{o2007caution}. 

\section{Results}


\subsection{RQ1: Psychosocial Effects of News Exposure}\label{rq:rq1}

Table~\ref{tab:lexicosemantics} presents an overview of the psychosocial changes observed between the matched \Tr{} and \Ct{} users following exposure to \news{} feed, spanning affective, behavioral, and cognitive outcomes, and Figure~\ref{fig:duration_effects} shows how these effects evolve over the next two months following treatment.
In summary, we note significant changes across several measures in both the directions of increase (RTE$>$1) and decrease (RTE$<$1) following \news{} feed exposure. We elaborate on these observations below:


\begin{figure*}[t]
    \centering
    \begin{subfigure}[b]{0.66\columnwidth}
        \centering
        \includegraphics[width=\textwidth]{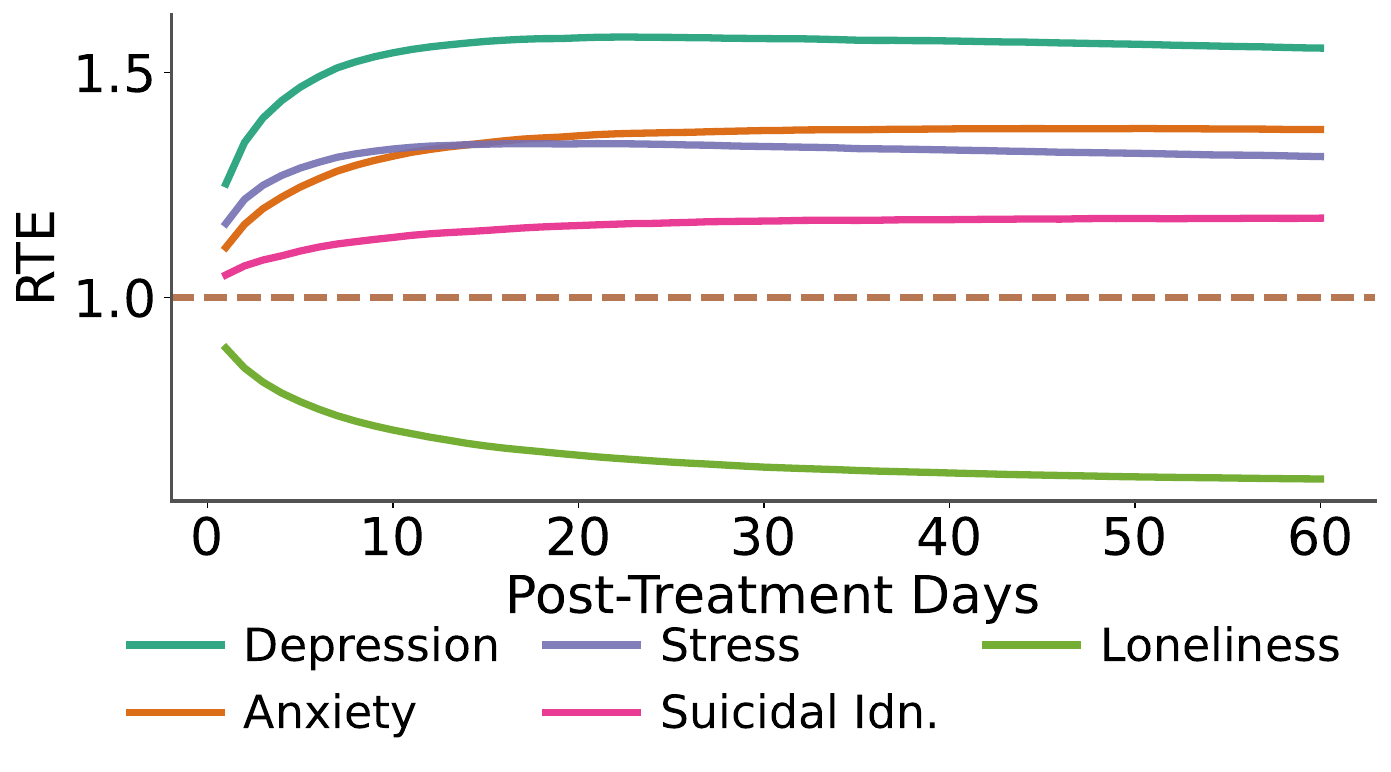}
        \caption{Affective Outcomes}
        \label{fig:duration_affective}
    \end{subfigure}
    \hfill
    \begin{subfigure}[b]{0.66\columnwidth}
        \centering
        \includegraphics[width=\textwidth]{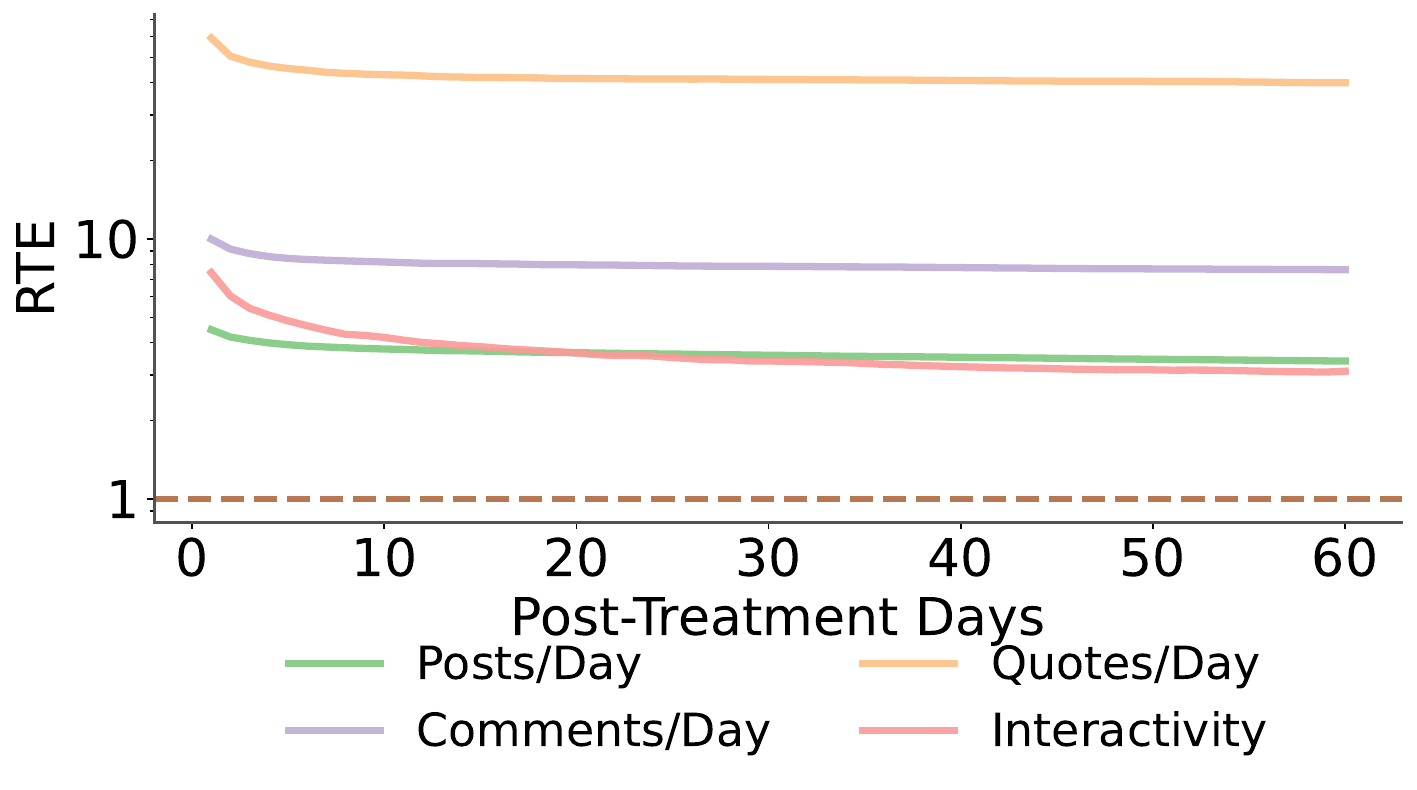}
        \caption{Behavioral Outcomes}
        \label{fig:duration_behavioral}
    \end{subfigure}
    \hfill
    \begin{subfigure}[b]{0.66\columnwidth}
        \centering
        \includegraphics[width=\textwidth]{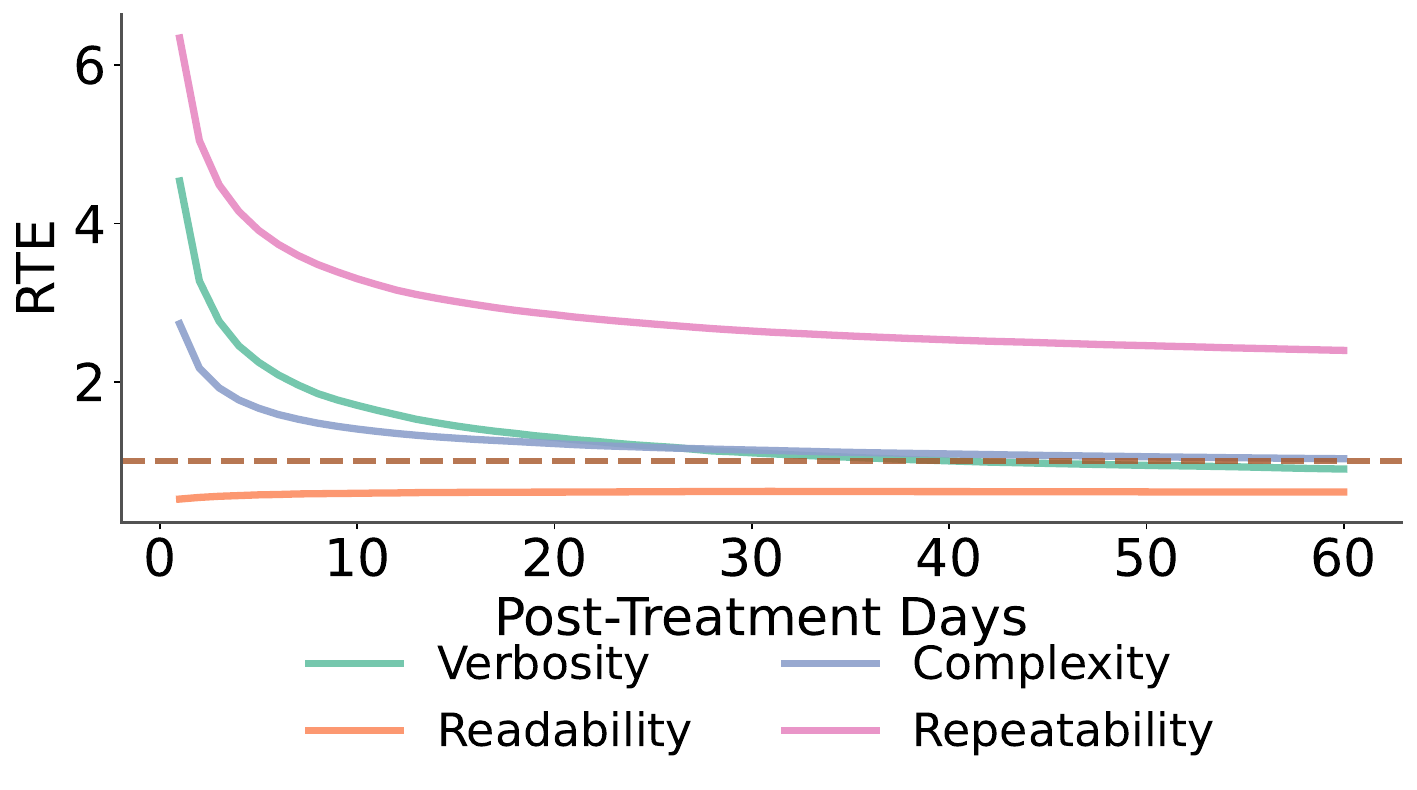}
        \caption{Cognitive Outcomes}
        \label{fig:duration_cognitive}
    \end{subfigure}
    
    \caption{Aggregated relative treatment effects (RTE) on psychosocial outcomes measured over two months following exposure.}
    \label{fig:duration_effects}
\end{figure*}

\subsubsection{Affective Outcomes}

The \Tr{} group showed significantly higher symptomatic mental health expressions, most notably 51\% higher depression (RTE=1.51), followed by anxiety (RTE=1.36) and stress (RTE=1.28).
Additionally, the \Tr{} group showed 56\% higher negative affect (RTE=1.56) and 63\% higher anger (RTE=1.63).
Further, Figure~\ref{fig:duration_affective} reveals that 
depression, anxiety, stress, and suicidal ideation expressions increase sharply among \Tr{} users immediately after \news{} feed exposure and then gradually attenuate before stabilizing.
The sharp post-exposure increases followed by gradual attenuation align with prior research on the models of stress and habituation, in which initial emotional arousal partially subsides while maintaining an elevated baseline~\cite{garfin2015cumulative}.

In contrast, \Tr{} users' loneliness expressions were 41\% lower (RTE=0.59), and loneliness expressions sharply dropped after treatment, and continued to decrease over the next two months (Figure~\ref{fig:duration_affective}).
The sustained decrease in loneliness suggests that news engagement may facilitate shared attention and social connection, echoing prior findings that participation in collective online discourse can reduce perceived isolation~\cite{boczkowski2018news,saha2017stress}.





\subsubsection{Behavioral Outcomes}
\Tr{} users show significantly higher behavioral outcomes than \Ct{} users. 
\Tr{} users show a significantly higher quantity of quoting behavior (0.78 vs. 0.02 quotes per day; RTE=39.56) and commenting behavior (2.04 vs. 0.28 comments per day; RTE=7.20).
Further, \Tr{} users show 220\% higher overall interactivity (RTE=3.20) than \Ct{} users.


Following first exposure to the \news{} feed, posting frequency, commenting, quoting, and overall interactivity rise sharply in the immediate post-exposure period and then plateau, indicating sustained but stable engagement levels over the subsequent 60 days (Figure~\ref{fig:duration_behavioral}). 
These patterns are consistent with an initial surge in participation that stabilizes over time, but RTE continues to remain above 1, indicating shared attention and collective discourse.
The sharp post-exposure increase followed by a plateau is consistent with theories of attention dynamics, in which salient informational stimuli trigger an initial surge in participation that stabilizes as norms and routines of engagement emerge~\cite{lehmann2012dynamical}. Again, the sustained elevation in interactivity suggests that news feeds may serve as focal points for ongoing social coordination and shared interpretation, rather than producing only short-lived bursts of activity~\cite{shamma2009tweet}.



\subsubsection{Cognitive Outcomes}
The \Tr{} users' posted 59\% more verbose content (18.44 vs. 11.57 words per sentence, RTE=1.59), but 123\% more repeatable content (0.43 vs. 0.19, RTE=2.23) than \Ct{} users
This indicates that \Tr{} wrote longer but more reused content.
Figure~\ref{fig:duration_cognitive} further reveals how verbosity, repeatability, and linguistic complexity increased immediately after first exposure to the \news{} feed and then gradually attenuated before stabilizing, while readability remained consistently lower over time. 

\begin{figure}[t]
    \centering
        \centering
        \includegraphics[width=0.95\columnwidth]{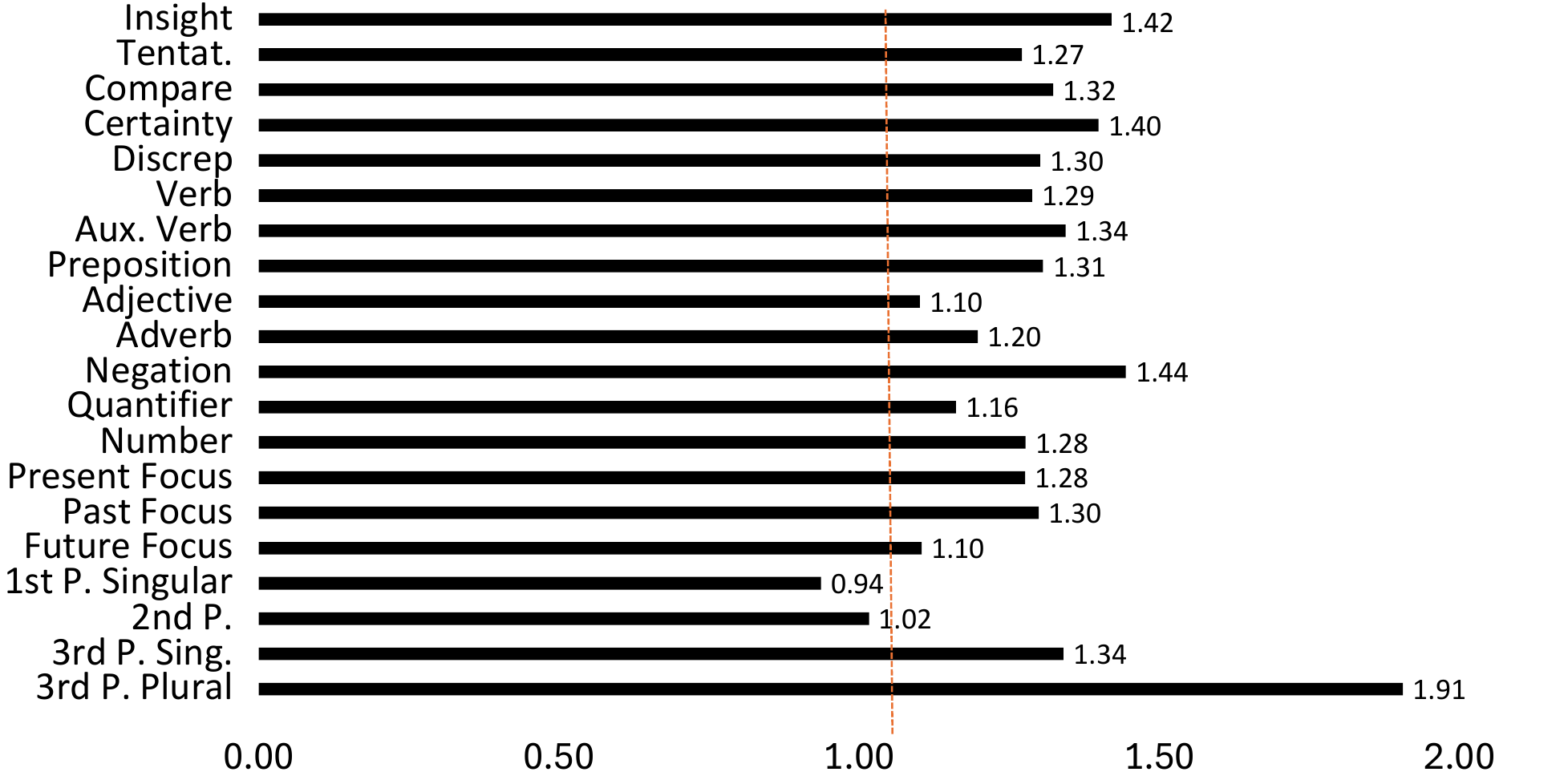}
        \label{fig:duration_mental}
       \caption{RTEs of LIWC cognitive and social attributes.}
    \label{fig:liwc_rte}
\end{figure}

\Tr{} users also showed a significantly larger quantity of psycholinguistic usage of cognitive (RTE=1.37) and social context (RTE=1.24).
In terms of function words, \Tr{} users exhibited higher overall usage (RTE=1.28), alongside increased use of personal pronouns (RTE=1.08), suggesting a shift toward more socially oriented language rather than purely individual expression. A more granular examination of Figure~\ref{fig:liwc_rte} reveals that \Tr{} users used fewer first-person singular pronouns (RTE=0.94), but substantially more third-person singular (RTE=1.34) and third-person plural pronouns (RTE=1.91). Together, these patterns are consistent with reduced self-focused attention and greater linguistic emphasis on others, aligning with prior psycholinguistic findings~\cite{pennebaker2003psychological,saha2024observer}.

\begin{table}[t!]
\centering
\small
\sffamily
\resizebox{\columnwidth}{!}{
\begin{tabular}{lrrrrr@{}l}
\setlength{\tabcolsep}{1pt}\\
\textbf{Metric} & \textbf{Tr.} & \textbf{Ct.} &\textbf{RTE}& \textbf{d} & \multicolumn{2}{c}{\textbf{t-test}}  \\ 
\toprule
\rowcollight \multicolumn{6}{c}{\textbf{Affective Outcomes}}\\
Depression & 0.74 & 0.49 & 1.51 & 0.86 & 174.03 & *** \\
Anxiety & 0.55 & 0.41 & 1.36 & 0.47 & 95.90 & *** \\
Stress & 0.66 & 0.51 & 1.28 & 0.46 & 93.27 & *** \\
\hdashline
Suicidal Ideation & 0.47 & 0.40 & 1.16 & 0.22 & 45.01 & *** \\
Loneliness & 0.20 & 0.34 & 0.59 & -0.70 & -141.90 & *** \\
LIWC: Anger & 0.77 & 0.47 & 1.63 & 0.21 & 38.61 & *** \\
\hdashline
LIWC: Neg. Affect & 1.86 & 1.19 & 1.56 & 0.31 & 56.45 & *** \\
LIWC: Pos. Affect & 4.97 & 4.71 & 1.05 & 0.05 & 9.45 & *** \\
\hdashline
LIWC: Sadness & 0.30 & 0.22 & 1.37 & 0.09 & 17.62 & *** \\

\hdashline
\rowcollight \multicolumn{6}{c}{\textbf{Behavioral Outcomes}}\\
Posting Frequency & 1.15 & 0.38 & 3.01 & 0.12 & 23.15 & *** \\
Quotes Frequency & 0.78 & 0.02 & 39.56 & 0.07 & 14.44 & *** \\
\hdashline
Comments Frequency & 2.04 & 0.28 & 7.20 & 0.17 & 34.77 & *** \\
\hdashline
Interactivity & 2.73 & 0.85 & 3.20 & 0.04 & 7.06 & *** \\

\hdashline
\rowcollight \multicolumn{6}{c}{\textbf{Cognitive Outcomes}}\\
Verbosity & 18.44 & 11.57 & 1.59 & 0.65 & 132.52 & *** \\
Readability & 8.83 & 14.86 & 0.59 & -0.36 & -64.32 & *** \\
Repeatability & 0.43 & 0.19 & 2.23 & 0.99 & 201.51 & *** \\
\hdashline
Complexity & 4.23 & 4.45 & 0.95 & -0.07 & -13.94 & *** \\
LIWC: Cog. Proc. & 8.08 & 5.91 & 1.37 & 0.39 & 71.66 & *** \\
LIWC: Social Context & 6.47 & 5.23 & 1.24 & 0.24 & 42.88 & *** \\
LIWC: Function Words & 36.26 & 28.37 & 1.28 & 0.39 & 72.92 & *** \\
LIWC: Personal Pronouns & 5.88 & 5.42 & 1.08 & 0.10 & 17.54 & *** \\

\end{tabular}}
\caption{Summary of comparisons of \Tr{} and \Ct{} datasets in terms of Relative Treatment Effect (RTE), effect size (Cohen's $d$), and paired $t$-test. Stars indicate statistical significance: *{$p$$<$0.05}, ** {$p$$<$0.01},  ***{$p$$<$0.001}.} 
\label{tab:lexicosemantics}
\end{table}

\subsection{RQ2: Degree of News Engagement and Psychosocial Outcomes} \label{rq:rq2}

\begin{table}[t]
\sffamily
\centering
\resizebox{\linewidth}{!}{
\setlength{\tabcolsep}{3pt}
\begin{tabular}{lrrrrrrr}
\textbf{Outcome} & \textbf{Bookmark} & \textbf{Comment} & \textbf{Like} & \textbf{Post} & \textbf{Quote} & \textbf{Repost} & $\mathbf{R^2}$\\
\midrule

\rowcollight \multicolumn{8}{c}{\textbf{Affective Outcomes}} \\
Depression                   & \cellcolor{sigthree}4.60 & \cellcolor{sigthree}0.30 & \cellcolor{sigthree}0.03 & -3E-3 & \cellcolor{sigthree}0.17 & \cellcolor{sigthree}0.07 & \cellcolor{sigthree} 0.250\\
Anxiety                      & \cellcolor{sigthree}5.60 & 0.02 & \cellcolor{sigone}8.7E-3 & -4.6E-3 & \cellcolor{sigthree}0.36 & \cellcolor{sigthree}-0.10 & \cellcolor{sigthree} 0.416\\
Stress                       & \cellcolor{sigthree}4.50 & \cellcolor{sigthree}0.19 & \cellcolor{sigtwo}0.01 & -4.9E-3 & \cellcolor{sigthree}0.20 & \cellcolor{sigthree}-0.09 & \cellcolor{sigthree} 0.439\\
Suicidal Ideation            & \cellcolor{sigone}-2.80 & \cellcolor{sigthree}-0.18 & 1.6E-3 & -2.1e-05 & \cellcolor{sigthree}-0.37 & \cellcolor{sigone}0.03 & \cellcolor{sigthree} 0.283\\
Loneliness                   & \cellcolor{sigthree}-4.40 & \cellcolor{sigthree}-0.15 & \cellcolor{sigthree}-0.02 & 3.8E-3 & \cellcolor{sigtwo}-0.03 & -0.01 & \cellcolor{sigthree} 0.412\\
LIWC: Anger                  & \cellcolor{sigtwo}-11.80 & \cellcolor{sigthree}0.88 & 3.9E-3 & 1.9e-05 & \cellcolor{sigthree}0.58 & \cellcolor{sigthree}-0.26 & \cellcolor{sigthree} 0.044\\
LIWC: Neg. Affect            & -11.60 & \cellcolor{sigthree}1.60 & 1.5E-3 & 8.2e-05 & \cellcolor{sigthree}0.95 & \cellcolor{sigthree}-0.53 & \cellcolor{sigthree} 0.109\\
LIWC: Pos. Affect            & \cellcolor{sigone}32.00 & 0.27 & 0.05 & 0.0002 & \cellcolor{sigthree}-1.40 & \cellcolor{sigthree}-0.90 & \cellcolor{sigthree} 0.229\\
LIWC: Sadness                & 0.0093 & 0.14 & -2.1E-4 & 2.5E-3 & -6E-3 & -0.07 & \cellcolor{sigthree} 0.016\\

\rowcollight \multicolumn{8}{c}{\textbf{Behavioral Outcomes}} \\
Posting Frequency            & -5.90 & \cellcolor{sigthree}4.60 & 0.04 & \cellcolor{sigthree}0.85 & \cellcolor{sigthree}5.30 & \cellcolor{sigone}0.86 & \cellcolor{sigthree} 0.011\\
Quotes Frequency             & -0.290 & -4.3E-3 & -0.14 & -0.08 & \cellcolor{sigthree}18.00 & 0.06 & \cellcolor{sigthree}0.007\\
Comments Frequency           & 33.70 & \cellcolor{sigthree}17.20 & \cellcolor{sigone}0.49 & -0.33 & \cellcolor{sigthree}18.80 & -0.65 & \cellcolor{sigthree}0.028\\
Interactivity                & \cellcolor{sigone}-573.50 & \cellcolor{sigone}-24.30 & -0.59 & -0.0022 & \cellcolor{sigthree}167.40 & \cellcolor{sigthree}37.90 & \cellcolor{sigthree}0.022\\

\rowcollight \multicolumn{8}{c}{\textbf{Cognitive Outcomes}} \\
Verbosity & \cellcolor{sigthree}-1.053 & \cellcolor{sigtwo}-2.80 & \cellcolor{sigtwo}-0.33 & 0.16 & \cellcolor{sigthree}6.50 & \cellcolor{sigthree}-1.50 & \cellcolor{sigthree}0.422\\
Readability  & -21.90 & \cellcolor{sigtwo}-1.70 & -0.06 & 0.0010 & \cellcolor{sigthree}6.70 & \cellcolor{sigthree}0.73 & \cellcolor{sigthree}0.520\\
Repeatability & \cellcolor{sigthree}4.70 & \cellcolor{sigthree}0.41 & \cellcolor{sigthree}0.04 & \cellcolor{sigthree}0.02 & \cellcolor{sigthree}0.58 & \cellcolor{sigthree}0.05 & \cellcolor{sigthree}0.325\\
Complexity  & -0.028 & 0.04 & 0.03 & 0.01 & \cellcolor{sigthree}0.86 & \cellcolor{sigthree}0.22 & \cellcolor{sigthree}0.273\\
LIWC: Cog. Proc. & \cellcolor{sigthree}53.40 & \cellcolor{sigthree}3.40 & 0.02 & -0.0014 & \cellcolor{sigthree}1.30 & \cellcolor{sigthree}-2.20 & \cellcolor{sigthree}0.417\\
LIWC: Social Context         & \cellcolor{sigtwo}37.90 & \cellcolor{sigthree}3.60 & 0.01 & -0.0004 & \cellcolor{sigthree}1.50 & \cellcolor{sigthree}-1.40 & \cellcolor{sigthree}0.280\\
LIWC: Function Words         & \cellcolor{sigthree}182.60 & \cellcolor{sigthree}11.00 & 0.14 & -0.33 & -0.16 & \cellcolor{sigthree}-8.50 & \cellcolor{sigthree}0.535\\
LIWC: Personal Pronouns & \cellcolor{sigone}0.262 & \cellcolor{sigthree}0.019 & 0.0005 & \cellcolor{sigone}-0.0014 & \cellcolor{sigthree}-0.024 & \cellcolor{sigthree}-0.016 & \cellcolor{sigthree}0.356\\
\end{tabular}
}
\caption{Linear regression estimates ($\beta \times 10^{-2}$) showing associations between engagement frequency (columns) and psychosocial outcomes (rows).
Shading indicates statistical significance: {\setlength{\fboxsep}{1pt}\colorbox{sigone}{$p$$<$0.05}}, {\setlength{\fboxsep}{1pt}\colorbox{sigtwo}{$p$$<$0.01}}, {\setlength{\fboxsep}{1pt}\colorbox{sigthree}{$p$$<$0.001}}.}
\label{tab:rq2_all}
\end{table}

For RQ2, we examine how the degree of news engagement is associated with psychosocial outcomes.  
For each outcome, we fit a regression model in which the dependent variable was the outcome's individual treatment effect (ITE), and the independent variables were the frequencies of various engagement types: bookmarking, commenting, liking, posting, quoting, and reposting.
Table~\ref{tab:rq2_all} summarizes the $\beta$ coefficients of the independent variables (columns) for each psychosocial outcome (rows). Across outcomes, several engagement measures show significant associations, suggesting that the degree of engagement meaningfully relates to psychosocial responses. This pattern is consistent with a ``dosage'' interpretation, in which greater levels of engagement correspond to stronger psychosocial effects.
 Notably, the frequency of bookmarking showed the strongest association, especially with affective outcomes,
 and quoting showed the strongest association with behavioral outcomes.
 We elaborate our observations below:

\subsubsection{Affective Outcomes}

For \textit{depression}, each additional bookmarked feed is associated with a positive coefficient ($\beta$=4.60), substantially larger than the effects of commenting ($\beta$=0.30), quoting ($\beta$=0.17), reposting ($\beta$=0.07), or liking ($\beta$=0.03).
A similar pattern is observed for \textit{anxiety} ($\beta$=5.60) and \textit{stress} ($\beta$=4.50), with bookmarking effects being ten times larger than the effects from commenting, quoting, and liking. 
Interestingly, reposting showed small but negative associations with both stress ($\beta$=-0.09) and anxiety ($\beta$=-0.10), indicating that each additional repost was associated with slightly decreased stress and anxiety. 

The pattern is different for \textit{suicidal ideation} and \textit{loneliness}. 
Quoting frequency showed the strongest negative association with suicidal ideation ($\beta$=-0.37), indicating that each additional quote was associated with decreased suicidal ideation. 
Commenting ($\beta$=-0.18) and bookmarking ($\beta$=-2.80) also showed negative associations, while reposting showed a small positive association ($\beta$=0.03). 

For \textit{loneliness}, bookmarking frequency showed the largest effect ($\beta$=-4.40), indicating that each additional bookmarked feed was associated with decreased loneliness. 
Commenting ($\beta$=-0.15), liking ($\beta$=-0.02), and quoting ($\beta$=-0.03) also showed negative associations. 
The negative coefficients indicate that increased news engagement, whether it is passive or active, corresponds with reduced loneliness, consistent with the findings in RQ1.

In addition, similar patterns emerge for LIWC-based affective expressions.
Bookmarking frequency showed strong negative associations with \textit{anger} ($\beta$=-11.80) and \textit{negative affect} ($\beta$=-11.60), while commenting and quoting showed positive associations with both anger ($\beta$=0.88 and $\beta$=0.58, respectively) and negative affect ($\beta$=1.60 and $\beta$=0.95).
\textit{Positive affect} exhibited a distinct pattern, with bookmarking showing a strong positive association ($\beta$=32.00), whereas quoting and reposting showed negative associations ($\beta$=-1.40 and $\beta$=-0.90).
Effects for \textit{sadness} were comparatively small and inconsistent across engagement types.

\subsubsection{Behavioral Outcomes}

For behavioral outcomes, quoting showed the strongest association with posting activity ($\beta$=5.30), indicating that higher levels of quoting were associated with increased posting frequency.
Commenting ($\beta$=4.60) and reposting ($\beta$=0.86) also showed positive associations with posting frequency.
Therefore, users who quoted or commented on news showed higher levels of overall platform participation, indicating a positive relationship between news engagement and posting activities.

For \textit{interactivity}, which is measured as the ratio of comments to posts, showed mixed associations across outcomes.
Quoting showed a strong positive association with interactivity ($\beta$=167.40), indicating substantially higher interactivity ratios.
Reposting ($\beta$=37.90) also showed positive associations.
Conversely, bookmarking ($\beta$=-573.50) and commenting ($\beta$=-24.30) showed strong negative associations, indicating that increased bookmarking or commenting frequency was associated with lower interactivity ratios.
The large negative coefficient for bookmarking indicates that users who primarily accumulate bookmarked news feeds show substantially lower interactive engagement, while those who actively share and quote news demonstrate higher interactivity ratios.

Consistent patterns are also observed for activity frequency measures.
\textit{Quoting frequency} shows the strongest positive association with quotes per day ($\beta$=18.00) and comments per day ($\beta$=18.80), while commenting frequency is also strongly associated with increased comments per day ($\beta$=17.20).
Bookmarking shows negative or negligible associations with posting and quoting frequency.

\subsubsection{Cognitive Outcomes}


For \textit{readability}, quoting \news{} content exhibited the largest positive association ($\beta$=6.70), suggesting that each additional quote corresponded to higher readability scores. Reposting also showed a smaller positive association ($\beta$=0.73), whereas commenting was negatively associated with readability ($\beta$=-1.70).
For \textit{complexity}, both quoting and reposting showed positive coefficients, indicating that users who engage in these behaviors tend to produce language with greater complexity and readability. 
This pattern may reflect the uptake of news-related linguistic structures or participation in more sophisticated discussions surrounding quoted or reposted content.


\textit{Verbosity} patterns varied distinctly by engagement type. 
For instance, posting ($\beta$=0.16) and quoting ($\beta$=6.50) were associated with longer content. 
Conversely, reposting showed a negative association with \textit{verbosity} ($\beta$=-1.50).
This pattern suggests that content creation (posting, quoting) encourages more elaborate expression, whereas simple redistribution (reposting) may not.

With respect to \textit{complexity}, quoting showed the greatest positive association ($\beta$=0.86), followed by reposting ($\beta$=0.22).
This suggests that greater engagement through quoting or reposting is associated with the use of more complex language, i.e., use of longer words.
Increased word length for these engagement types suggests users adopt more sophisticated vocabulary.

For \textit{repeatability}, all engagement types showed significant positive associations.
Bookmarking showed the largest effect ($\beta$=4.70), indicating that each additional feed bookmarking was associated with more repetitive language; users recycled the same phrases more frequently. Quoting ($\beta$=0.58), commenting ($\beta$=0.41), reposting ($\beta$=0.05), liking ($\beta$=0.04), and posting ($\beta$=0.02) also showed positive associations. The positive associations indicate that increased news engagement corresponds with more repetitive language, suggesting users might recycle phrases and concepts.

\section{Discussion and Conclusion}\label{section:discussion}

This work examined the psychosocial effects of news exposure on social media using a large-scale quasi-experimental design, revealing systematic yet nuanced patterns in how routine news engagement shapes users' affective, behavioral, and cognitive wellbeing. 
Rather than producing uniformly negative or positive outcomes, news engagement was associated with a set of trade-offs that depended on how users engaged with news content. In this section, we discuss the theoretical implications of these findings for media-effects research, followed by practical and design implications for platforms, tools, and interventions.

\subsubsection{Theoretical Implications}
Our findings extend existing theories of media and news effects in critical ways. \textit{First}, our work extends theories of news effects beyond crisis-centric frameworks by revealing the psychosocial consequences of routine news engagement. 
Prior research has largely focused on acute events such as pandemics, mass violence, or natural disasters, where distress follows temporally bounded exposure~\cite{thompson2019media}. 
In contrast, our results show that everyday news engagement embedded in routine social media use can significantly alter psychosocial wellbeing. This suggests that theories of news effects must account not only for episodic shocks but also for the cumulative psychological burden of persistent informational environments.

\textit{Second}, our findings highlight engagement heterogeneity in affecting wellbeing.
Research often conceptualizes exposure primarily in terms of volume~\cite{garfin2020novel,thompson2019media}, implicitly assuming that greater engagement leads to stronger effects regardless of how users engage. 
Our findings challenge this assumption by showing that different engagement forms are associated with varying outcomes.
In particular, bookmarking was associated with significantly greater increases in depression, anxiety, and stress than active forms such as commenting or quoting.
These differences were not marginal; per-unit effects of consumption exceeded those of engagement through liking, commenting, and reposting by an order of magnitude. 
Theoretically, this calls for a shift away from treating news exposure as a uniform construct and toward models that distinguish not only how users consume news, but also how they participate in the news ecosystem.


\textit{Third}, this work refines theoretical accounts of the relationship between news engagement on social media and wellbeing by revealing coexisting affective costs and social benefits. 
This pattern complicates narratives that frame news consumption as either uniformly harmful or beneficial. 
Instead, it suggests a dual-process dynamic in which news engagement facilitates social connection and collective sense-making, even as it imposes emotional strain on individuals. 
Theoretically, this aligns with perspectives that emphasize the social functions of news in coordinating attention and fostering shared understanding, while underscoring the psychological costs of sustained exposure.



\subsubsection{Methodological, Practical, and Design Implications}
Our findings not only suggest interesting design levers for mitigating the psychosocial costs of news engagement, but also highlight a fundamental tension between user wellbeing and platform incentives. 
Multiple patterns associated with adverse psychosocial outcomes---such as the accumulation of multiple news feeds and sustained passive consumption---are precisely those that maximize time spent on platform and repeated exposure. 
Interventions that meaningfully reduce these patterns would likely run counter to platforms' core economic objectives. 
Indeed, if large social media platforms were primarily oriented toward optimizing users' psychological wellbeing, contemporary news feed ecosystems would likely be designed very differently.

This misalignment raises critical questions about the feasibility of platform-led solutions. 
While it is common to frame wellbeing-oriented interventions as matters of responsible design, our results suggest that such interventions may be structurally disincentivized within engagement-driven business models. 
As a consequence, the burden of mitigating the psychosocial effects of news exposure may fall disproportionately on users themselves, as well as on researchers, designers, and civic technologists operating outside platform governance structures.

In this context, user-facing and post-hoc interventions become particularly salient. Tools that operate independent of platform ranking and recommendation systems---such as browser extensions or third-party overlays---could enable users to monitor cumulative news exposure, differentiate passive from active engagement, and selectively reorganize or filter their feeds in ways that better align with their emotional and cognitive capacities.
Importantly, such tools do not require changes to platform algorithms or incentives, making them more feasible in practice.

There is a growing need for more personalized approaches to feed curation, including algorithms that enable users to actively customize their feeds according to individual preferences. Recent work on designing teachable feeds~\cite{choi2025designing} and supporting user-driven curation~\cite{malki2025bonsai} on \bsky{} present a promising direction for future research and system design.

Beyond reducing exposure, our findings also suggest opportunities to reshape how users engage with news. 
Design interventions that encourage commenting, discussion, or contextualization---rather than endless feed accumulation---may help preserve the social benefits of news engagement, while mitigating its emotional costs. Shifting engagement away from purely consumptive patterns and toward more interactive or reflective participation may offer a pathway to balance psychosocial trade-offs without requiring users to disengage from news entirely.

Our findings also highlight the potential role of community-driven and alternative platforms, where governance and incentive structures differ from those of dominant commercial platforms. 
These platforms may be better positioned to experiment with designs that limit exposure, which could affect users' mental health, and promote more contextualized engagement with news (and other content). 
Such experimental platform designs can serve as critical testbeds for translating insights about psychosocial wellbeing into design practices that are difficult to implement within profit-driven ecosystems.

More broadly, these implications suggest that addressing the psychological consequences of news exposure requires us to move beyond the assumption that platforms will voluntarily internalize wellbeing costs. 
Rather, meaningful and alternative progress may depend on empowering users with greater control over their informational environments and supporting external, post-hoc mechanisms for moderation and curation. 
From this perspective, the role of research is not only to identify harmful patterns of engagement, but also to inform the design of tools and systems that enable individuals and communities to reclaim agency in increasingly attention-driven news ecosystems.

Finally, our work underscores the value of quasi-experimental methods for measuring psychosocial outcomes in new platforms. 
By observing users' early engagement trajectories and distinguishing between participation types, researchers can better isolate the effects of specific design features and behaviors. Methodologically, this approach complements survey and laboratory studies by capturing longitudinal, large-scale dynamics in naturalistic settings.

\subsection{Ethical Considerations}

This research employs computational and data-driven approaches to study the psychosocial effects of routine news exposure on social media users. Given the sensitivity of mental health–related outcomes and the potential vulnerability of affected populations, the ethical implications of this work are multifaceted and warrant careful consideration. 
In particular, computational models used to infer symptomatic mental health expressions or affective states from social media language can be misused or misappropriated in ways that may cause harm. 
For example, such models could be leveraged for targeted advertising, commercial profiling, surveillance, or insurance-related decision-making, rather than for research or wellbeing-oriented purposes.
We caution against interpreting these models as diagnostic tools or as providing clinical assessments of mental health conditions. Our findings reflect population-level patterns inferred from behavioral and linguistic expressions and should not be used to draw conclusions about individual users' mental health status. Any deployment of tools inspired by this work would require substantial additional safeguards, including clear limitations on use, transparency, and oversight by domain experts.
Further, social media expressions are shaped by cultural, social, and contextual factors, and mental health experiences and coping mechanisms vary widely across individuals and communities. 
Applying these methods without sensitivity to such differences risks reinforcing stereotypes or biases about particular groups or forms of expression. Researchers and practitioners should therefore take care to avoid overgeneralization and to consider how differences in language use, cultural norms, and platform practices may influence observed patterns.
Overall, this work is intended to inform empirical understanding of psychosocial dynamics in online news environments, not to enable monitoring, profiling, or intervention at the individual level. We emphasize the importance of ethical reflection, responsible use, and contextualized interpretation when studying mental health–related phenomena using computational approaches.

\subsubsection{Limitations and Future Directions}

Our study has limitations, which also suggest interesting future directions. 
We focus on how users engage with news rather than on the substantive topics or narrative frames of the news itself. 
While our results show that engagement modality and intensity shape psychosocial outcomes, different types of news---such as political conflict, public health crises, or local community events---may have distinct psychological consequences even under similar engagement patterns. 
Future work should examine how topic-specific news exposure interacts with forms of engagement to influence mental health.

Although our quasi-experimental design strengthens causal inference relative to purely observational approaches, it does not establish \textit{true causality}. Moreover, our findings are not clinical in nature and are inferred from users' behavioral and linguistic expressions on \bsky{}, rather than from validated mental health assessments. This inspires future work with natural experiments and mixed-methods designs, and validated mental health surveys to further strengthen both causal inference and clinical interpretability.


Finally, our analysis is limited to a single platform, \bsky{}, raising questions about generalizability. Studying news (and other online) engagement across platforms is increasingly difficult due to restricted data access, which often limits external researchers to a narrow set of observable interactions, such as commenting, while obscuring passive consumption behaviors that may be most consequential for mental health. 
As a result, some forms of research can realistically be conducted only within platforms. Addressing these constraints may therefore require new research models, including data donation initiatives, privacy-preserving data sharing, and stronger industry–academia partnerships, to enable broader and more representative analyses while maintaining user privacy.


\fontsize{9pt}{8pt} {\selectfont
\bibliography{0paper}}

\begin{thebibliography}{82}
\providecommand{\natexlab}[1]{#1}

\bibitem[{Bakshy, Messing, and Adamic(2015)}]{bakshy2015exposure}
Bakshy, E.; Messing, S.; and Adamic, L.~A. 2015.
\newblock Exposure to ideologically diverse news and opinion on Facebook.
\newblock \emph{Science}.

\bibitem[{Boczkowski, Mitchelstein, and Matassi(2018)}]{boczkowski2018news}
Boczkowski, P.~J.; Mitchelstein, E.; and Matassi, M. 2018.
\newblock “News comes across when I’m in a moment of leisure”: Understanding the practices of incidental news consumption on social media.
\newblock \emph{New media \& society}, 20(10): 3523--3539.

\bibitem[{Boukes and Vliegenthart(2017)}]{boukes2017news}
Boukes, M.; and Vliegenthart, R. 2017.
\newblock News consumption and its unpleasant side effect.
\newblock \emph{Journal of Media Psychology}.

\bibitem[{Breckler(1984)}]{breckler1984empirical}
Breckler, S.~J. 1984.
\newblock Empirical validation of affect, behavior, and cognition as distinct components of attitude.
\newblock \emph{Journal of personality and social psychology}, 47(6): 1191.

\bibitem[{Burnap, Colombo, and Scourfield(2015)}]{burnap2015machine}
Burnap, P.; Colombo, W.; and Scourfield, J. 2015.
\newblock Machine classification and analysis of suicide-related communication on twitter.
\newblock In \emph{Proc. ACM conference on hypertext \& social media}.

\bibitem[{Chan et~al.(2026)Chan, Choi, Saha, and Chandrasekharan}]{chan2025examining}
Chan, J.; Choi, F.; Saha, K.; and Chandrasekharan, E. 2026.
\newblock Examining Algorithmic Curation on Social Media: An Empirical Audit of Reddit's r/popular Feed.
\newblock In \emph{Proceedings of the International AAAI Conference on Web and Social Media}.

\bibitem[{Chandrasekharan et~al.(2017)Chandrasekharan, Pavalanathan, Srinivasan, Glynn, Eisenstein, and Gilbert}]{chandrasekharan2017you}
Chandrasekharan, E.; Pavalanathan, U.; Srinivasan, A.; Glynn, A.; Eisenstein, J.; and Gilbert, E. 2017.
\newblock You can't stay here: The efficacy of reddit's 2015 ban examined through hate speech.
\newblock \emph{Proceedings of the ACM on human-computer interaction}, 1(CSCW): 1--22.

\bibitem[{Choi and Chandrasekharan(2025)}]{choi2025designing}
Choi, F.; and Chandrasekharan, E. 2025.
\newblock Designing Usable Controls for Customizable Social Media Feeds.
\newblock \emph{arXiv preprint arXiv:2509.19615}.

\bibitem[{Chowdhury et~al.(2021)Chowdhury, Saha, Hasan, Saha, and Mueen}]{chowdhury2021examining}
Chowdhury, F.~A.; Saha, D.; Hasan, M.~R.; Saha, K.; and Mueen, A. 2021.
\newblock Examining factors associated with twitter account suspension following the 2020 us presidential election.
\newblock In \emph{Proceedings of the 2021 IEEE/ACM international conference on advances in social networks analysis and mining}, 607--612.

\bibitem[{Cohn, Mehl, and Pennebaker(2004)}]{cohn2004linguistic}
Cohn, M.~A.; Mehl, M.~R.; and Pennebaker, J.~W. 2004.
\newblock Linguistic markers of psychological change surrounding September 11, 2001.
\newblock \emph{Psychol. Sci.}

\bibitem[{Coppersmith, Dredze, and Harman(2014)}]{coppersmith2014quantifying}
Coppersmith, G.; Dredze, M.; and Harman, C. 2014.
\newblock Quantifying mental health signals in twitter.
\newblock In \emph{Proc. ACL CLCP Workshop}.

\bibitem[{Culotta(2014)}]{culotta2014estimating}
Culotta, A. 2014.
\newblock Estimating county health statistics with Twitter.

\bibitem[{De~Choudhury and De(2014)}]{de2014mental}
De~Choudhury, M.; and De, S. 2014.
\newblock Mental health discourse on reddit: Self-disclosure, social support, and anonymity.
\newblock In \emph{ICWSM}.

\bibitem[{De~Choudhury et~al.(2013)De~Choudhury, Gamon, Counts, and Horvitz}]{de2013predicting}
De~Choudhury, M.; Gamon, M.; Counts, S.; and Horvitz, E. 2013.
\newblock Predicting depression via social media.
\newblock In \emph{ICWSM}.

\bibitem[{De~Choudhury and K{\i}c{\i}man(2017)}]{de2017language}
De~Choudhury, M.; and K{\i}c{\i}man, E. 2017.
\newblock The language of social support in social media and its effect on suicidal ideation risk.
\newblock In \emph{ICWSM}.

\bibitem[{De~Choudhury et~al.(2016)De~Choudhury, Kiciman, Dredze, Coppersmith, and Kumar}]{de2016discovering}
De~Choudhury, M.; Kiciman, E.; Dredze, M.; Coppersmith, G.; and Kumar, M. 2016.
\newblock Discovering shifts to suicidal ideation from mental health content in social media.
\newblock In \emph{CHI}.

\bibitem[{De~Choudhury, Monroy-Hernandez, and Mark(2014)}]{de2014narco}
De~Choudhury, M.; Monroy-Hernandez, A.; and Mark, G. 2014.
\newblock Narco emotions: affect and desensitization in social media during the mexican drug war.
\newblock In \emph{CHI}, 3563--3572. ACM.

\bibitem[{Ernala et~al.(2019)Ernala, Birnbaum, Candan, Rizvi, Sterling, Kane, and De~Choudhury}]{ernala2019methodological}
Ernala, S.~K.; Birnbaum, M.~L.; Candan, K.~A.; Rizvi, A.~F.; Sterling, W.~A.; Kane, J.~M.; and De~Choudhury, M. 2019.
\newblock Methodological gaps in predicting mental health states from social media: triangulating diagnostic signals.
\newblock In \emph{Proceedings of the 2019 CHI Conference on Human Factors in Computing Systems}, 1--16.

\bibitem[{Ernala et~al.(2017)Ernala, Rizvi, Birnbaum, Kane, and De~Choudhury}]{ernala2017linguistic}
Ernala, S.~K.; Rizvi, A.~F.; Birnbaum, M.~L.; Kane, J.~M.; and De~Choudhury, M. 2017.
\newblock Linguistic markers indicating therapeutic outcomes of social media disclosures of schizophrenia.
\newblock \emph{PACM HCI}, (CSCW).

\bibitem[{Failla and Rossetti(2024)}]{failla2024m}
Failla, A.; and Rossetti, G. 2024.
\newblock “I’m in the Bluesky Tonight”: Insights from a year worth of social data.
\newblock \emph{PloS one}.

\bibitem[{Garfin, Holman, and Silver(2015)}]{garfin2015cumulative}
Garfin, D.~R.; Holman, E.~A.; and Silver, R.~C. 2015.
\newblock Cumulative exposure to prior collective trauma and acute stress responses to the Boston marathon bombings.
\newblock \emph{Psychological science}.

\bibitem[{Garfin, Silver, and Holman(2020)}]{garfin2020novel}
Garfin, D.~R.; Silver, R.~C.; and Holman, E.~A. 2020.
\newblock The novel coronavirus (COVID-2019) outbreak: Amplification of public health consequences by media exposure.
\newblock \emph{Health psychology}.

\bibitem[{Gerbner et~al.(2006)Gerbner, Gross, Morgan, and Signorielli}]{gerbner1980mainstreaming}
Gerbner, G.; Gross, L.; Morgan, M.; and Signorielli, N. 2006.
\newblock The ``Mainstreaming'' of America: Violence Profile No. 11.
\newblock \emph{Journal of Communication}, 30(3): 10--29.

\bibitem[{Goyal, Lambert, and Chandrasekharan(2025)}]{goyal2025language}
Goyal, A.; Lambert, C.; and Chandrasekharan, E. 2025.
\newblock The language of approval: Identifying the drivers of positive feedback online.
\newblock \emph{arXiv preprint arXiv:2509.10370}.

\bibitem[{Grimmelmann(2015)}]{grimmelmann2015law}
Grimmelmann, J. 2015.
\newblock The law and ethics of experiments on social media users.
\newblock \emph{Colo. Tech. LJ}, 13: 219.

\bibitem[{Guess et~al.(2020)Guess, Lerner, Lyons, Montgomery, Nyhan, Reifler, and Sircar}]{guess2020digital}
Guess, A.~M.; Lerner, M.; Lyons, B.; Montgomery, J.~M.; Nyhan, B.; Reifler, J.; and Sircar, N. 2020.
\newblock A digital media literacy intervention increases discernment between mainstream and false news in the United States and India.
\newblock \emph{Proceedings of the National Academy of Sciences}, 117(27): 15536--15545.

\bibitem[{Guntuku et~al.(2019)Guntuku, Ramsay, Merchant, and Ungar}]{guntuku2019language}
Guntuku, S.~C.; Ramsay, J.~R.; Merchant, R.~M.; and Ungar, L.~H. 2019.
\newblock Language of ADHD in adults on social media.
\newblock \emph{Journal of attention disorders}.

\bibitem[{Hermida(2010)}]{hermida2010twittering}
Hermida, A. 2010.
\newblock Twittering the news: The emergence of ambient journalism.
\newblock \emph{Journalism practice}, 4(3): 297--308.

\bibitem[{Huff(2022)}]{huff2022media}
Huff, C. 2022.
\newblock Media overload is hurting our mental health. Here are ways to manage headline stress.
\newblock \emph{Monit. Psychol}, 53: 20.

\bibitem[{Hwang et~al.(2021)Hwang, Borah, Shah, and Brauer}]{hwang2021relationship}
Hwang, J.; Borah, P.; Shah, D.; and Brauer, M. 2021.
\newblock The relationship among COVID-19 information seeking, news media use, and emotional distress at the onset of the pandemic.
\newblock \emph{International Journal of Environmental Research and Public Health}, 18(24): 13198.

\bibitem[{Jha et~al.(2021)Jha, Awasthi, Kumar, Kumar, and Sethi}]{jha2021learning}
Jha, I.~P.; Awasthi, R.; Kumar, A.; Kumar, V.; and Sethi, T. 2021.
\newblock Learning the mental health impact of COVID-19 in the United States with explainable artificial intelligence: observational study.
\newblock \emph{JMIR mental health}.

\bibitem[{Jhaver, Rathi, and Saha(2024)}]{jhaver2024bystanders}
Jhaver, S.; Rathi, H.; and Saha, K. 2024.
\newblock Bystanders of online moderation: Examining the effects of witnessing post-removal explanations.
\newblock In \emph{Proceedings of the 2024 CHI Conference on Human Factors in Computing Systems}, 1--9.

\bibitem[{K\i{}c\i{}man, Counts, and Gasser(2018)}]{kiciman2018using}
K\i{}c\i{}man, E.; Counts, S.; and Gasser, M. 2018.
\newblock Using Longitudinal Social Media Analysis to Understand the Effects of Early College Alcohol Use.
\newblock In \emph{ICWSM}, 171--180.

\bibitem[{Kolden et~al.(2011)Kolden, Klein, Wang, and Austin}]{kolden2011congruence}
Kolden, G.~G.; Klein, M.~H.; Wang, C.-C.; and Austin, S.~B. 2011.
\newblock Congruence/genuineness.
\newblock \emph{Psychotherapy}, 48(1): 65.

\bibitem[{Lambert, Saha, and Chandrasekharan(2025)}]{lambert2025does}
Lambert, C.; Saha, K.; and Chandrasekharan, E. 2025.
\newblock Does Positive Reinforcement Work?: A Quasi-Experimental Study of the Effects of Positive Feedback on Reddit.
\newblock In \emph{Proceedings of the 2025 CHI Conference on Human Factors in Computing Systems}, 1--16.

\bibitem[{Lehmann et~al.(2012)Lehmann, Gon{\c{c}}alves, Ramasco, and Cattuto}]{lehmann2012dynamical}
Lehmann, J.; Gon{\c{c}}alves, B.; Ramasco, J.~J.; and Cattuto, C. 2012.
\newblock Dynamical classes of collective attention in twitter.
\newblock In \emph{Proceedings of the 21st international conference on World Wide Web}, 251--260.

\bibitem[{Lin et~al.(2014)Lin, Jia, Guo, Xue, Li, Huang, Cai, and Feng}]{lin2014user}
Lin, H.; Jia, J.; Guo, Q.; Xue, Y.; Li, Q.; Huang, J.; Cai, L.; and Feng, L. 2014.
\newblock User-level psychological stress detection from social media using deep neural network.
\newblock In \emph{ACM Multimedia}. ACM.

\bibitem[{Lin and Margolin(2014)}]{lin2014ripple}
Lin, Y.-R.; and Margolin, D. 2014.
\newblock The ripple of fear, sympathy and solidarity during the Boston bombings.
\newblock \emph{EPJ Data Science}.

\bibitem[{Malki et~al.(2025)Malki, Qu{\'e}r{\'e}, Monroy-Hern{\'a}ndez, and Ribeiro}]{malki2025bonsai}
Malki, O.~E.; Qu{\'e}r{\'e}, M. A.~L.; Monroy-Hern{\'a}ndez, A.; and Ribeiro, M.~H. 2025.
\newblock Bonsai: Intentional and personalized social media feeds.
\newblock \emph{arXiv preprint arXiv:2509.10776}.

\bibitem[{Mark et~al.(2012)Mark, Bagdouri, Palen, Martin, Al-Ani, and Anderson}]{mark2012blogs}
Mark, G.; Bagdouri, M.; Palen, L.; Martin, J.; Al-Ani, B.; and Anderson, K. 2012.
\newblock Blogs as a collective war diary.
\newblock In \emph{CSCW}.

\bibitem[{McCombs and Shaw(1972)}]{mccombs1972agenda}
McCombs, M.~E.; and Shaw, D.~L. 1972.
\newblock The agenda-setting function of mass media.
\newblock \emph{Public opinion quarterly}, 36(2): 176--187.

\bibitem[{McLaughlin, Gotlieb, and Mills(2023)}]{mclaughlin2023caught}
McLaughlin, B.; Gotlieb, M.~R.; and Mills, D.~J. 2023.
\newblock Caught in a dangerous world: Problematic news consumption and its relationship to mental and physical ill-being.
\newblock \emph{Health communication}.

\bibitem[{Messing and Westwood(2014)}]{messing2014selective}
Messing, S.; and Westwood, S.~J. 2014.
\newblock Selective exposure in the age of social media: Endorsements trump partisan source affiliation when selecting news online.
\newblock \emph{Communication research}.

\bibitem[{Moreno et~al.(2013)Moreno, Goniu, Moreno, and Diekema}]{moreno2013ethics}
Moreno, M.~A.; Goniu, N.; Moreno, P.~S.; and Diekema, D. 2013.
\newblock Ethics of social media research: Common concerns and practical considerations.
\newblock \emph{Cyberpsychology, behavior, and social networking}.

\bibitem[{Newman et~al.(2025)Newman, Ross~Arguedas, Robertson, Nielsen, and Fletcher}]{newman2025digital}
Newman, N.; Ross~Arguedas, A.; Robertson, C.~T.; Nielsen, R.~K.; and Fletcher, R. 2025.
\newblock \emph{Digital news report 2025}.
\newblock Reuters Institute for the study of Journalism.

\bibitem[{Nogara et~al.(2025)Nogara, Sahneh, DeVerna, Liu, Luceri, Menczer, Pierri, and Giordano}]{nogara2025}
Nogara, G.; Sahneh, E.~S.; DeVerna, M.~R.; Liu, N.; Luceri, L.; Menczer, F.; Pierri, F.; and Giordano, S. 2025.
\newblock A longitudinal analysis of misinformation, polarization and toxicity on Bluesky after its public launch.
\newblock arXiv:2505.02317.

\bibitem[{O\'brien(2007)}]{o2007caution}
O\'brien, R.~M. 2007.
\newblock A caution regarding rules of thumb for variance inflation factors.
\newblock \emph{Quality \& quantity}, 41(5): 673--690.

\bibitem[{Olteanu et~al.(2019)Olteanu, Castillo, Diaz, and Kiciman}]{olteanu2019social}
Olteanu, A.; Castillo, C.; Diaz, F.; and Kiciman, E. 2019.
\newblock Social data: Biases, methodological pitfalls, and ethical boundaries.
\newblock \emph{Frontiers in Big Data}, 2: 13.

\bibitem[{Palen(2008)}]{palen2008online}
Palen, L. 2008.
\newblock Online social media in crisis events.
\newblock \emph{Educause Quarterly}.

\bibitem[{Park, Conway et~al.(2018)}]{park2018harnessing}
Park, A.; Conway, M.; et~al. 2018.
\newblock Harnessing reddit to understand the written-communication challenges experienced by individuals with mental health disorders: analysis of texts from mental health communities.
\newblock \emph{Journal of medical Internet research}, 20(4): e8219.

\bibitem[{Pennebaker et~al.(2015)Pennebaker, Boyd, Jordan, and Blackburn}]{pennebaker2015development}
Pennebaker, J.~W.; Boyd, R.~L.; Jordan, K.; and Blackburn, K. 2015.
\newblock The Development and Psychometric Properties of LIWC2015.
\newblock Technical report, University of Texas at Austin.

\bibitem[{Pennebaker, Francis, and Booth(2001)}]{pennebaker2001linguistic}
Pennebaker, J.~W.; Francis, M.~E.; and Booth, R.~J. 2001.
\newblock Linguistic inquiry and word count: LIWC 2001.
\newblock \emph{Mahway: Lawrence Erlbaum Associates}, 71: 2001.

\bibitem[{Pennebaker, Mehl, and Niederhoffer(2003)}]{pennebaker2003psychological}
Pennebaker, J.~W.; Mehl, M.~R.; and Niederhoffer, K.~G. 2003.
\newblock Psychological aspects of natural language use: Our words, our selves.
\newblock \emph{Annual review of psychology}, 54(1): 547--577.

\bibitem[{Pennebaker and Stone(2003)}]{pennebaker2003words}
Pennebaker, J.~W.; and Stone, L.~D. 2003.
\newblock Words of wisdom: language use over the life span.
\newblock \emph{Journal of personality and social psychology}, 85(2): 291.

\bibitem[{{Pew Research Center}(2025)}]{pewsocialmedia2025}
{Pew Research Center}. 2025.
\newblock Social Media and News Fact Sheet.
\newblock Survey of U.S. adults conducted Aug. 18-24, 2025.

\bibitem[{Rubin(2005)}]{rubin2005causal}
Rubin, D.~B. 2005.
\newblock Causal inference using potential outcomes: Design, modeling, decisions.
\newblock \emph{J. Am. Stat. Assoc.}

\bibitem[{Saha, Chandrasekharan, and De~Choudhury(2019)}]{saha2019prevalence}
Saha, K.; Chandrasekharan, E.; and De~Choudhury, M. 2019.
\newblock Prevalence and psychological effects of hateful speech in online college communities.
\newblock In \emph{WebSci}.

\bibitem[{Saha and De~Choudhury(2017)}]{saha2017stress}
Saha, K.; and De~Choudhury, M. 2017.
\newblock Modeling stress with social media around incidents of gun violence on college campuses.
\newblock \emph{PACM HCI}, (CSCW).

\bibitem[{Saha et~al.(2024)Saha, Gupta, Mark, Kiciman, and De~Choudhury}]{saha2024observer}
Saha, K.; Gupta, P.; Mark, G.; Kiciman, E.; and De~Choudhury, M. 2024.
\newblock Observer Effect in Social Media Use.
\newblock In \emph{CHI}.

\bibitem[{Saha et~al.(2021)Saha, Liu, Vincent, Chowdhury, Neves, Shah, and Bos}]{saha2021advertiming}
Saha, K.; Liu, Y.; Vincent, N.; Chowdhury, F.~A.; Neves, L.; Shah, N.; and Bos, M.~W. 2021.
\newblock Advertiming matters: Examining user ad consumption for effective ad allocations on social media.
\newblock In \emph{CHI}.

\bibitem[{Saha and Sharma(2020)}]{saha2020causal}
Saha, K.; and Sharma, A. 2020.
\newblock Causal Factors of Effective Psychosocial Outcomes in Online Mental Health Communities.
\newblock In \emph{ICWSM}.

\bibitem[{Saha et~al.(2019)Saha, Sugar, Torous, Abrahao, K{\i}c{\i}man, and De~Choudhury}]{saha2019social}
Saha, K.; Sugar, B.; Torous, J.; Abrahao, B.; K{\i}c{\i}man, E.; and De~Choudhury, M. 2019.
\newblock A Social Media Study on the Effects of Psychiatric Medication Use.
\newblock In \emph{ICWSM}.

\bibitem[{Saha et~al.(2020)Saha, Torous, Caine, and De~Choudhury}]{saha2020psychosocial}
Saha, K.; Torous, J.; Caine, E.~D.; and De~Choudhury, M. 2020.
\newblock Psychosocial effects of the COVID-19 pandemic: large-scale quasi-experimental study on social media.
\newblock \emph{JMIR}.

\bibitem[{Saha, Weber, and De~Choudhury(2018)}]{saha2018social}
Saha, K.; Weber, I.; and De~Choudhury, M. 2018.
\newblock A Social Media Based Examination of the Effects of Counseling Recommendations After Student Deaths on College Campuses.
\newblock In \emph{ICWSM}.

\bibitem[{Saha et~al.(2022)Saha, Yousuf, Boyd, Pennebaker, and De~Choudhury}]{saha2022social}
Saha, K.; Yousuf, A.; Boyd, R.~L.; Pennebaker, J.~W.; and De~Choudhury, M. 2022.
\newblock Social media discussions predict mental health consultations on college campuses.
\newblock \emph{Scientific reports}.

\bibitem[{Shamma, Kennedy, and Churchill(2009)}]{shamma2009tweet}
Shamma, D.~A.; Kennedy, L.; and Churchill, E.~F. 2009.
\newblock Tweet the debates: understanding community annotation of uncollected sources.
\newblock In \emph{Proceedings of the first SIGMM workshop on Social media}, 3--10.

\bibitem[{Shearer and Mitchell(2021)}]{mitchellpew}
Shearer, E.; and Mitchell, A. 2021.
\newblock News Use Across Social Media Platforms in 2020.
\newblock Pew Research Center.

\bibitem[{Silver et~al.(2013)Silver, Holman, Andersen, Poulin, McIntosh, and Gil-Rivas}]{silver2013mental}
Silver, R.~C.; Holman, E.~A.; Andersen, J.~P.; Poulin, M.; McIntosh, D.~N.; and Gil-Rivas, V. 2013.
\newblock Mental-and physical-health effects of acute exposure to media images of the September 11, 2001, attacks and the Iraq War.
\newblock \emph{Psychological science}, 24(9): 1623--1634.

\bibitem[{Small(2024)}]{small2024protocols}
Small, D.~S. 2024.
\newblock Protocols for observational studies: Methods and open problems.
\newblock \emph{Statistical Science}, 39(4): 519--554.

\bibitem[{Soroka and McAdams(2015)}]{soroka2015news}
Soroka, S.; and McAdams, S. 2015.
\newblock News, politics, and negativity.
\newblock \emph{Political communication}, 32(1): 1--22.

\bibitem[{Starbird et~al.(2010)Starbird, Palen, Hughes, and Vieweg}]{starbird2010chatter}
Starbird, K.; Palen, L.; Hughes, A.~L.; and Vieweg, S. 2010.
\newblock Chatter on the red: what hazards threat reveals about the social life of microblogged information.
\newblock In \emph{CSCW}, 241--250. ACM.

\bibitem[{Strasser, Sumner, and Meyer(2022)}]{strasser2022covid}
Strasser, M.~A.; Sumner, P.~J.; and Meyer, D. 2022.
\newblock COVID-19 news consumption and distress in young people: A systematic review.
\newblock \emph{Journal of affective disorders}, 300: 481--491.

\bibitem[{Tausczik and Pennebaker(2010)}]{tausczik2010psychological}
Tausczik, Y.~R.; and Pennebaker, J.~W. 2010.
\newblock The psychological meaning of words: LIWC and computerized text analysis methods.
\newblock \emph{Journal of language and social psychology}, 29(1): 24--54.

\bibitem[{Thompson et~al.(2019)Thompson, Jones, Holman, and Silver}]{thompson2019media}
Thompson, R.~R.; Jones, N.~M.; Holman, E.~A.; and Silver, R.~C. 2019.
\newblock Media exposure to mass violence events can fuel a cycle of distress.
\newblock \emph{Science advances}, 5(4): eaav3502.

\bibitem[{Thorson and Wells(2016)}]{thorson2016curated}
Thorson, K.; and Wells, C. 2016.
\newblock Curated flows: A framework for mapping media exposure in the digital age.
\newblock \emph{Communication theory}.

\bibitem[{Tourangeau, Rips, and Rasinski(2000)}]{tourangeau2000psychology}
Tourangeau, R.; Rips, L.~J.; and Rasinski, K. 2000.
\newblock \emph{The psychology of survey response}.

\bibitem[{Valdes, Eisenstein, and De~Choudhury(2015)}]{valdes2015psychological}
Valdes, J. M.~D.; Eisenstein, J.; and De~Choudhury, M. 2015.
\newblock Psychological effects of urban crime gleaned from social media.
\newblock In \emph{ICWSM}.

\bibitem[{Vowels et~al.(2023)Vowels, Vowels, Carnelley, Millings, and Gibson-Miller}]{vowels2023toward}
Vowels, L.~M.; Vowels, M.~J.; Carnelley, K.~B.; Millings, A.; and Gibson-Miller, J. 2023.
\newblock Toward a causal link between attachment styles and mental health during the COVID-19 pandemic.
\newblock \emph{Br. J. Clin. Psychol.}

\bibitem[{Xie, Brand, and Jann(2012)}]{xie2012estimating}
Xie, Y.; Brand, J.~E.; and Jann, B. 2012.
\newblock Estimating heterogeneous treatment effects with observational data.
\newblock \emph{Sociol. Methodol.}

\bibitem[{Yuan et~al.(2023)Yuan, Saha, Keller, Isomets{\"a}, and Aledavood}]{yuan2023mental}
Yuan, Y.; Saha, K.; Keller, B.; Isomets{\"a}, E.~T.; and Aledavood, T. 2023.
\newblock Mental health coping stories on social media: a causal-inference study of Papageno effect.
\newblock In \emph{TheWebConf}.

\bibitem[{Yuan et~al.(2026)Yuan, Zhang, Aledavood, Zhang, and Saha}]{yuan2025mental}
Yuan, Y.; Zhang, J.; Aledavood, T.; Zhang, R.; and Saha, K. 2026.
\newblock Mental Health Impacts of AI Companions: Triangulating Social Media Quasi-Experiments, User Perspectives, and Relational Theory.
\newblock In \emph{Proceedings of the 2026 CHI Conference on Human Factors in Computing Systems}.

\bibitem[{Zhu et~al.(2009)Zhu, Zou, Rosset, Hastie et~al.}]{zhu2009multi}
Zhu, J.; Zou, H.; Rosset, S.; Hastie, T.; et~al. 2009.
\newblock Multi-class adaboost.
\newblock \emph{Statistics and its Interface}, 2(3): 349--360.

\end{thebibliography}






\end{document}